\documentstyle[preprint,aps,amsfonts]{revtex}   
\tighten   
\draft   
\widetext   
\input epsf   
\preprint{CLNS-99/1627, ITP-SB-99-37}   
\bigskip   
\bigskip   
\begin{document}   
\title{Anomaly Cancelations in Orientifolds with Quantized B Flux}   
\medskip   
\author{Alex Buchel$^{1}$\footnote{E-mail:   
buchel@mail.lns.cornell.edu\\
\hbox{$\;\;$} Address after September 1, 1999: ITP - Kohn Hall,
University of California
Santa Barbara,\\ \hbox{$\;\;$} CA 93106-4030.},   
Gary Shiu$^{2}$\footnote{E-mail:   
shiu@insti.physics.sunysb.edu} and S.-H. Henry Tye$^{3}$\footnote{E-mail:   
tye@mail.lns.cornell.edu}}   
\bigskip   
\address{$^{1,3}$Newman Laboratory of Nuclear Studies, Cornell University,    
Ithaca, NY 14853\\   
$^2$C.N. Yang Institute for Theoretical Physics, \\ 
State University of   
New York,  
Stony Brook, NY 11794}   
\bigskip   
\medskip   
\maketitle

\def\vt{\vartheta}   
\def\preal{{\rm Re\,}}   
\def\pim{{\rm Im\,}}   
\def\ds{\displaystyle}   
\def\yzero{\smash{\hbox{$y\kern-4pt\raise1pt\hbox{${}^\circ$}$}}}   
\def\p{\partial}   
\def\a{\alpha}   
\def\b{\beta}   
\def\g{\gamma}   
\def\d{\delta}   
\def\Om{\Omega}   
\def\om{\omega}   
\def\th{\theta}   
\def\vt{\vartheta}   
\def\vphi{\varphi}   
\def\-{\hphantom{-}}   
\def\ov{\overline}   
\def\s2{\frac{1}{\sqrt2}}   
\def\wh{\widehat}   
\def\wt{\widetilde}   
\def\oh{\frac{1}{2}}   
\def\beq{\begin{equation}}   
\def\eeq{\end{equation}}   
\def\beqa{\begin{eqnarray}}   
\def\eeqa{\end{eqnarray}}   
\def\tr{{\rm tr \,}}   
\def\Tr{{\rm Tr \,}}   
\def\diag{{\rm diag \,}}   
\def\IF{\relax{\rm I\kern-.18em F}}   
\def\II{\relax{\rm I\kern-.18em I}}   
\def\IP{\relax{\rm I\kern-.18em P}}   
\def\IC{\relax\hbox{\kern.25em$\inbar\kern-.3em{\rm C}$}}   
\def\IR{\relax{\rm I\kern-.18em R}}   
\def\hm{\relax{n_H}}   
\def\vac{|0 \rangle}   
\def\vm{\relax{n_V}}   
\def\cc{{\cal C}}   
\def\ck{{\cal K}}   
\def\ci{{\cal I}}   
\def\cu{{\cal U}}   
\def\cg{{\cal G}}   
\def\cn{{\cal N}}   
\def\cam{{\cal M}}   
\def\cp{{\cal P}}   
\def\ct{{\cal T}}   
\def\cv{{\cal V}}   
\def\cz{{\cal Z}}   
\def\ch{{\cal H}}   
\def\cf{{\cal F}}   
\def\tv{\tilde v}   
\def\Dsl{\,\raise.15ex\hbox{/}\mkern-13.5mu D} 
\def\IZ{Z\kern-.4em  Z}   
\def\id{{\rm I}}   
\def\bP{{\bf\rm P}}   
\def\bF{{\bf\rm F}}   
\def\ep{{\epsilon}}   
\def\mbz{{\mathbb Z}}  
\def\bz{{\bf Z}}   
\def\d{{\delta}}

\begin{abstract}   
   
{}We consider anomaly cancelations in Type IIB orientifolds on
$T^4/\mbz_N$ with quantized NS-NS sector background B-flux. For a rank
$b$ B-flux on $T^4$ ($b$ is always even) and when $N$ is even, the
cancelation requires a $2^{b/2}$ multiplicity of states in the 59-open
string sector. We identify the twisted sector R-R scalars and tensor
multiplets which are involved in the Green-Schwarz mechanism.  We give
more details of the construction of these models and argue that
consistency with the $2^{b/2}$ multiplicity of 59-sector states
requires a modification of the relation between the open string 1-loop
channel modulus and the closed string tree channel modulus in the
59-cylinder amplitudes.
   
\end{abstract}   
\pacs{}   
   
\section{Introduction}   
\label{introduction6d}   
   
Probably the most attractive feature of string theory is that it
incorporates both quantum gravity and gauge interactions in a
consistent framework. A consistent string model should therefore have
the appropriate particle content and interactions such that all
apparent field-theoretic gauge and/or gravitational anomalies are
absent.  In four dimensional string models, there can be $U(1)_A$
gauge symmetries which are pseudo-anomalous. However, the
Green-Schwarz mechanism \cite{gs} ensures that counterterms are
generated from the exchange of pseudoscalar fields and any
field-theoretic $U(1)_A$ triangle anomaly is canceled by the
corresponding non-trivial $U(1)_A$ transformation of the
pseudoscalars.
   
The analysis of anomalous $U(1)$ have mainly be focused on
perturbative heterotic string theory \cite{hetgs} where many
semi-realistic string models can be easily constructed.  In recent
years, however, tremendous progress has been made in constructing
four-dimensional ${\cal N}=1$ Type I string models \cite{typeI}.
Moreover, it was recently suggested that if the Standard Model
particles are localized on a set of $p$-branes where $3<p<9$ whereas
gravity lives in higher dimensional spacetime, then the string scale
can be as low as a few TeV \cite{lyk,TeV,ST} .  Type I string theory
provides a natural framework to realize this ``brane world'' scenario
\cite{TeV,ST,BW}.  In view of these developments, it is therefore  
important to carry out similar analysis for Type I string vacua.
   
Anomaly cancelations in some four-dimensional ${\cal N}=1$ Type I
string models were studied in some recent papers \cite{iru9808}.
Anomaly cancelations in $D=6$ Type IIB $\mbz_N$ orientifolds without
the B-flux background were studied in \cite{sagnotti,6authors,ss99}.  
The analysis involves techniques in computing couplings of the 
twisted moduli with the gauge fields which were also recently studied in \cite{dudas}.
These
simple examples illustrate some new features in Type I string vacua
that are distinct from that of the perturbative heterotic string.
First, the R-R scalars which participate in the Green-Schwarz
mechanism come from the twisted sectors which can couple differently
to different gauge factors. Therefore, in contrast to perturbative
heterotic string, there can be more than one anomalous $U(1)$ in these
Type I vacua. Moreover, the Fayet-Iliopoulos term already appears at
tree-level and is dependent on the blowing-up modes of the orbifolds.
   
In this paper, we consider orientifold models with a non-vanishing
NS-NS sector B-field background \cite{Bij,w9712,kst9803}.  Although
the fluctuations of the NS-NS B-field are projected out of the
orientifold spectrum (since it is odd under worldsheet parity
reversal), a quantized vacuum expectation value of this B-field is
allowed.  Furthermore, the presence of this B-field gives rise to some
novel features.  In particular, the rank of the gauge group is reduced
by $2^{b/2}$ \cite{Bij,w9712,kst9803} where $b$ is the rank of the
background B-field.  Moreover, in models where there are both $D9$ and
$D5$-branes, the $59$ sector states come with a multiplicity of
$2^{b/2}$ \cite{kst9803}.  These reduced rank orientifold models are
dual to the CHL strings \cite{CHL} in heterotic string theory.  For
models with both $D9$ and $D5$-branes, the corresponding heterotic
duals contain NS fivebranes. Therefore, the analysis of these
orientifolds may shed some lights on the non-perturbative properties
of the CHL strings.

Following the work of Sen \cite{sen}, we also expect a close relation
between orientifolds and F theory vacua. In Ref \cite{bershad}, some
eight-dimensional F theory vacua with non-zero background B-flux were
studied.  Naively, the presence of this NS-NS sector B-field
background is incompatible with the $SL(2,\mbz)$ symmetry (the
S-duality) of Type IIB theory.  However, the presence of this B-flux
freezes 8 of the moduli of the elliptic $K3$, leaving only a
10-dimensional subspace.  The monodromy group on this subspace is
reduced from $SL(2,\mbz)$ to the congruence group $\Gamma_0(2)$ which
is the largest subgroup of $SL(2,\mathbb{Z})$ that keeps the B-flux
invariant.  In this paper, we consider models with both $D9$ and
$D5$-branes. They can be viewed as generalizations of F theory
compactifications of this kind to lower dimensions when there are more
than one type of 7-branes.
   
On the other hand, this class of models is closely related to the
setup considered in
\cite{ncg}.  According to \cite{ncg}, the positions of the branes  
cease to commute in the presence of a B-field background. It is
possible that our results here can be understood from the point of
view of noncommutative geometry.
   
Type IIB orientifolds with quantized background B-field are also
interesting from the phenomenological point of view.  In the absence
of B-field, the residual gauge symmetries are typically too large for
the models to be phenomenologically interesting.  Since the rank of
the gauge group is reduced by $2^{b/2}$ in the presence of B-field, it
is possible to construct string models containing the Standard Model
with fewer additional gauge symmetries.  In fact, the three-family
Pati-Salam like model in Ref. \cite{ST} was constructed by turning on
a non-zero background B-field with $b=2$ in a $\mbz_6$ orientifold.
There are both $D9$ and $D5$-branes in this model. Under T-duality,
they become 2 sets of $D5$-branes: $D5_1$ and $D5_2$ whose
intersection is our four-dimensional non-compact spacetime.  The
Standard Model $SU(3)$ lives on the $D5_1$-branes and the three chiral
families of fermions in the Standard Model come from open strings with
at least one end attached to the $D5_1$-branes.  One of the families
comes from $5_15_1$ open strings whereas the other two families are
$5_15_2$ open string states.  The fact that there are two families in
the $5_15_2$ sector depends crucially on the multiplicity of
$2^{b/2}=2$ in the $5_15_2$ sector.
   
Since the main new features in models with non-zero B-field that we
discuss here already appear in six dimensions, we will focus our
attention to six-dimensional models in this paper.  Moreover, as we
will see, anomaly cancelations in six dimensions provide rather
stringent constraints on the consistencies of these models.  In
particular, we will show that the multiplicity of $2^{b/2}$ in the
$59$ sector is crucial for the cancelation of the leading anomalies
({\em i.e.}, the $\tr F^4$ and $\tr R^4$ terms). Moreover, the
remaining anomalies are properly factorized which can then be canceled
by the Green-Schwarz mechanism.  The analysis of four-dimensional
${\cal N}=1$ orientifold models with background B-field will appear in
a separate publication \cite{bst}.
   
This paper is organized as follows. In Section \ref{6d}, we describe
in detail the construction of some six-dimensional orientifold models
with non-zero background B-field considered in Ref.\cite{kst9803}.  In
Section \ref{anomaly}, we discuss anomaly cancelations in these
models. We end with some discussion in Section \ref{discussion}. Some
of the technical details are relegated to the appendices.
   
\section{Six Dimensional Examples}   
\label{6d}   
   
In this section, we discuss in detail the construction of
six-dimensional Type IIB orientifolds with background B-flux
\cite{kst9803}.  In particular, we consider the orbifold limits of K3:  
$T^4/\mbz_N$ where $N=2,3,4,6$.\footnote{These orientifolds without
background B-flux have been constructed in \cite{gp,gj9606}.}  Since
we are interested in models with both $D9$ and $D5$-branes, we only
consider the cases $N=2,4,6$.  However, turning on the NS-NS
antisymmetric two-form background seems to render the $T^4/\mbz_6$
Type IIB perturbative orientifold inconsistent\footnote{$T^4/\mbz_6$
Type IIB orientifold with rank two B-flux does not contain states $\bf
(4,4;1,1)$ and $\bf (1,1;4,4)$ given in Table II of \cite{kst9803}.
As a result, the $\tr R^4$ and $\tr F^4$ anomalies do not
cancel. However, it is consistent \cite{bst} to turn on B-field in the
$T^6/\mbz_6$ orientifold in Ref.\cite{ST} since the corresponding
$\mbz_3$ twist is different.}, so we will concentrate on the $\mbz_2$
and $\mbz_4$ models.  Type I vacua on smooth K3 with non-zero
$B$-field have been studied in \cite{SS}.
       
Toroidal compactification of Type IIB string theory on a four
dimensional torus $T^4$ gives rise to a six dimensional model with
$\cn=4$ supersymmetry. By gauging the world-sheet parity $\Om$
$(X_L\leftrightarrow X_R)$ of Type IIB strings reduces by half the
number of supersymmetries. One can further reduce the number of
supersymmetries to $\cn=1$ by orbifolding. Specifically, the $\mbz_N$
orbifold action is realized by powers of the twist generator $\theta $
($\theta^N=1$) which can be written in the form
\begin{equation}   
\theta= \exp (2i\pi (v_1J_{67}+v_2J_{89})),   
\label{twist}   
\end{equation}   
where $J_{mn}$ are $SO(4)$ Cartan generators and
$v\equiv(v_1,v_2)={\frac{1}N }(1,-1)$ represents the twist.  In terms
of the complex bosonic coordinates $Y_1 = X_6+iX_7$, $Y_2 = X_8+iX_9$
that parametrize the torus, $\th$ acts diagonally as
\begin{equation}   
\theta^k Y_i  = {\rm e}^{2i\pi kv_i}Y_i.   
\label{thdi}   
\end{equation}   
Similarly, we define complex fermionic fields $\psi^i$ as   
$\psi^1= \psi^6+i\psi^7$ and $\psi^2= \psi^8+i\psi^9$.   
   
To derive the massless spectra of the orientifolds, we will work in
light-cone gauge. For example, in the closed untwisted sector the NS
massless states are $\psi_{-\oh}^\mu \vac$ which are invariant under
$\th$, and $\psi_{-\oh}^i \vac$ which transforms as
\beq   
\th^k  \psi_{-\oh}^i \vac =  {\rm e}^{2i\pi kv_i} \psi_{-\oh}^i \vac.   
\label{thpsi}   
\eeq   
Complex conjugates $\psi_{-\oh}^{\bar{i}}$ transform with a phase   
${\rm e}^{-2i\pi kv_i}$.   
The untwisted massless Ramond states are of the form   
$| s_0 s_1 s_2 s_3 \rangle$   
with $s_0, s_i = \pm \oh$. To implement the GSO projection    
we retain only the states with $s_0+s_1+s_2+s_3=0,\ {\rm mod\ 2}$.   
These states transform as   
\beq   
\theta^k |s_0 s_1 s_2 s_3 \rangle = {\rm e}^{2i\pi kv\cdot s}   
|s_0 s_1 s_2 s_3 \rangle.   
\label{twistR}   
\eeq   
   
The close string spectrum is obtained by retaining only those states
in the untwisted sector which are invariant under the orientifold
group action and by including twisted sector states. This will be
discussed in more detail in Sec.~\ref{closedsector}.
   
Although Type IIB theory is a theory of closed strings, the
orientifold projection requires both closed and open string sectors
for consistency.  The Klein bottle amplitude (which is present due to
the orientifold projection) in the closed string sector generically
gives rise to tadpole divergences. These divergences are canceled by
the new contributions from the open string sector \cite{pcai}.
Alternatively, orientifold fixed planes are sources for the
Ramond-Ramond (R-R) $(p+1)$-forms.  Charge cancelation can be
generically achieved by including the right number of D$p$-branes,
carrying opposite charge with respect to these forms \cite{polchin}.
The endpoints of an open string are labeled by $a$ and $b$ which lie
on a $D_p$ and $D_q$-brane respectively (the corresponding excitations
are $pq$ sector states).  The models we discuss here have both D9 and
D5-branes.  The open string sector of the models will be discussed in
Sec.~\ref{opensector}.

\subsection{Open String Sector}   
\label{opensector}   
   
The open string spectrum of Type IIB orientifolds on $T^4/\mbz_N$ are
determined by the type and the number of D-branes necessary to cancel
the tadpole divergences in the closed string Klein bottle amplitude.
The worldsheet parity element $\Omega$ of the orientifold group
produces the tadpole which must be canceled by the 9-branes.  For even
$N$, the closed string sector tadpole from $\Omega \th^{N/2}$
orientifold element is canceled by introducing $5$-branes, whose
worldvolume spans the uncompactified $D=6$ space-time.
   
Let $\Psi$ be a world-sheet excitation and $a,b$ represent Chan-Paton
indices associated with the string endpoints on D$p$-brane and
D$q$-brane ($p,q=\{5,9\}$).  These Chan-Paton indices must be
contracted with a hermitian matrix $\lambda_{ab}$.  The action of the
group elements on the open string state $|\Psi, ab \rangle $ is given
by \cite{gp}
\beqa   
\th^k:&&\ |\Psi, ab \rangle \rightarrow     
(\gamma _{k,p})_{aa'} |\th^k.\Psi,a'b'   
\rangle (\gamma _{k,q})_{b'b}^{-1},\cr   
\Omega\th^k:&&\ |\Psi, ab \rangle \rightarrow     
(\gamma _{\Omega_k,p})_{aa'} |\Omega\th^k.\Psi,b'a'   
\rangle (\gamma _{\Omega_k,q})_{b'b}^{-1},   
\label{groupaction}   
\eeqa   
where $\g_{k,p}$ and $\g_{\Om_k,p}$ are unitary matrices associated
with $\th^k$.  In order to be consistent with group multiplication
(\ref{groupaction}), we can choose
\beq   
\g_{\Om_k,p} = \g_{k,p}\cdot \g_{\Om,p},   
\label{gamom}   
\eeq   
and    
\beq   
\g_{k,p}=\g_{1,p}^k.   
\label{gamk}   
\eeq    
The absence of pure gauge twists requires $\g_{0,p} = 1$  \cite{gp}.    
Because $\th^N=1$,    
\beq   
\gamma_{1,p}^N=\pm1.   
\label{gton}   
\eeq   
Otherwise $\g_{N,p}$ would be a pure gauge twist.   
Similarly, from $(\Omega \th^k)^2=\th^{2k}$,   
\beq   
\g_{\Om_k,p}=\pm \g_{2k,p} \, \g^T_{\Om_k,p}.   
\label{orcon1}   
\eeq   
Finally, using eqs.~(\ref{gamom}), (\ref{gamk}), (\ref{orcon1}) and the   
unitarity of the $\g$ matrices we obtain   
\beq   
\g_{k,p}^* = \pm \g_{\Om,p}^* \, \g_{k,p} \g_{\Om,p}.   
\label{afam}   
\eeq   
Since different types of branes are present it is also necessary to
consider the action of $(\Omega \th^k)^2$ on $pq$ states.  In
Ref.\cite{gp}, it was argued that $\Om^2 =-1$ on $95$ states. This
implies that in eq.~(\ref{orcon1}) there are opposite signs for 9 and
5-branes.  Since a simultaneous $T$-duality on $X_{6,7,8,9}$
interchanges 9 and 5- branes along with $\Om_0$ and $\Om_{N/2}$,
\beq   
\g_{\Om_0,p}=\pm\g^T_{\Om_0,p}\qquad \Rightarrow\qquad    
\g_{\Om_{N/2},p}=\mp\g^T_{\Om_{N/2},p},   
\label{tduality}   
\eeq    
where the various signs are now correlated. From    
eqs.~(\ref{orcon1})-(\ref{tduality}) it further follows that   
\beq   
\g_{1,p}^N=-1.   
\label{gton1}   
\eeq   
for both 9 and 5- branes.

Consider first the 99 open string states. The massless NS states   
include gauge bosons    
$\psi_{-\oh}^\mu|0,ab\rangle \, \lambda_{ab}^{(0)}$   
and matter scalars $\psi_{-\oh}^i |0,ab \rangle \, \lambda_{ab}^{(i)}$.   
The Chan-Paton matrices must be such that the full states are   
invariant under the action of the orientifold group. Hence,   
\beqa   
\lambda^{(0)} & = & \gamma_{1,9}\lambda^{(0)}{\gamma^{-1}_{1,9}}  \   
\quad\quad\quad ; \quad \quad   
\lambda^{(0)} = -\gamma_{\Omega,9} {\lambda^{(0)}}^T {\gamma^{-1}_{\Omega,9}},
\nonumber \\[0.3ex]   
\lambda^{(i)} & = & {\rm e}^{2\pi i v_i}\g_{1,9}\lambda^{(i)}\g^{-1}_{1,9}   
\quad\quad ; \quad \quad   
\lambda ^{(i)}  =  -\g_{\Omega,9}{\lambda^{(i)}}^T\g^{-1}_{\Omega,9}.   
\label{cons99}   
\eeqa   
   
The massless NS states of the 55 open strings also include gauge bosons   
$\psi_{-\oh}^\mu|0,ab\rangle \, \lambda_{ab}^{(0)}$   
and scalars $\psi_{-\oh}^i |0,ab \rangle \, \lambda_{ab}^{(i)}$.   
We consider  models where all $5$-branes sit at the origin    
of the orientifold. The Chan-Paton matrices then satisfy   
\beqa   
\lambda^{(0)} & = & \gamma_{1,5}\lambda^{(0)}{\gamma^{-1}_{1,5}} \   
\quad\quad\quad ; \quad \quad   
\lambda^{(0)} = -\gamma_{\Omega,5} {\lambda^{(0)}}^T 
{\gamma^{-1}_{\Omega,5}},   
\nonumber \\[0.3ex]   
\lambda^{(i)}  & = &e^{2\pi i v_i}\g_{1,5} \lambda^{(i)} {\g^{-1}_{1,5}}   
\quad\quad ; \quad \quad   
\lambda^{(i)}  =  \g_{\Omega,5} {\lambda^{(i)}}^T \g^{-1}_{\Omega,5}.   
\label{cons55}   
\eeqa   
The sign change in the $\Om$ projection is due to the DD boundary   
conditions on the  directions transverse to the $5$-branes.

Finally, consider $59$ open string states.  For $5$-branes, the
compact coordinates obey mixed DN boundary conditions and have
expansions with half-integer modded creation operators. By world-sheet
supersymmetry their fermionic partners in the NS sector are integer
modded. Their zero modes form a representation of a Clifford algebra
and can be labeled as $|s_i, s_j\rangle$, $i,j= \{2,3\}$, with $s_i,
s_j = \pm \oh$. The GSO projection further imposes $s_i=s_j$. Since
$|s_i, s_j\rangle$ are invariant under $\theta$, the orientifold
projection on these states implies
\begin{equation}   
\lambda = \g_{1,5} \lambda  {\g^{-1}_{1,9}}.   
\label{cons59}   
\end{equation}   
Notice that here the index $a$ ($b$) lies on a 5-brane (9-brane).   
$\Omega $ relates $59$  with $95$ sectors and does not impose extra   
constraints on $\lambda$.   
   
The perturbative spectrum of Type IIB theory contains two massless
antisymmetric tensor fields: one coming from the NS-NS sector, and
another coming from the R-R sector. Under the world-sheet parity
reversal, the NS-NS two-form $B_{\mu \nu}$ is projected out, while the
R-R two-form $B_{\mu \nu}^{\prime}$ is kept.  Although the
fluctuations of $B_{ij}$ (the components of $B_{\mu \nu}$ in the
compactified dimensions) are projected out of the perturbative
unoriented closed string spectrum, a quantized vacuum expectation
value of $B_{ij}$ is allowed. To see this, consider the left- and
right-moving momenta in the $4$ dimensions compactified on a torus
$T^{4}$:
\begin{equation}\label{momenta}   
P_{L,R}={\tilde{e}^i \over R_i} (n_i - b_{ij}m^j)     
\pm  {e_i m^i R_i\over \a^\prime}~,   
\label{bmomentum}   
\end{equation}   
where $m_i$ and $n^i$ are integers, $e_i$ are constant vielbeins such
that $e_i \cdot e_j=G_{ij}$ is the constant background metric on
$T^{4}$, $R_i$ are compactification radii of the $T^4$, and $e_i \cdot
\tilde{e}^{j}={\delta_i}^j$. In (\ref{bmomentum}) $b_{ij}=B_{ij} R_i   
R_j/\a^\prime$.  Note that the components of $b_{ij}$ are defined up
to a shift $b_{ij} \rightarrow b_{ij} + 1$ (which can be absorbed by
redefining $n_i$). With this normalization, only the values $b_{ij}=0$
and $1/2$ are invariant under $\Omega$, and hence the vacuum
expectation values of $b_{ij}$ are quantized.  Let $b=rank(B_{ij})$
($i,j\in \{6,7,8,9\}$) be the rank of NS-NS antisymmetric tensor, it
is clear that $b\in 2\mbz$.
   
The unitary matrices $\g_{k,p}$ and $\g_{\Omega_k,p}$ must be chosen
so that NS-NS tadpoles from the M\"obius strip (MS) and the cylinder
(C) of the open string sector cancel the NS-NS tadpole of the closed
string sector coming from the Klein bottle (KB).  The specific form of
$\g_{k,p}$ and $\g_{\Omega_k,p}$, and so the massless open string
sector of the orientifold, is sensitive to the NS-NS antisymmetric
tensor background.
   
Consider first the case of zero B-flux, $b=0$.  The 1-loop vacuum
amplitudes of the models are given in Appendix~\ref{1loopb0}.  The
$t\to 0$ divergences of the KB, C and MS (c.f.,
Eqs.(\ref{kamp}),(\ref{camp}) and (\ref{mamp})) produce tadpoles for
the untwisted sector R-R 6- and 10-form as well as the twisted sector
R-R 6-form.  Relating the tree channel modulus $\ell$ to the KB, MS
and C loop moduli through
\beq   
{\rm KB:}\ t_{\ck}={1\over 4\ell},\qquad  {\rm MS:}\ t_{\cam}={1\over 8\ell}, 
\qquad  {\rm C:}\ t_{\cc}={1\over 2\ell},   
\label{tvsl}   
\eeq   
all the tadpoles vanish provided\footnote{We consider models    
with all 5-branes at the origin.}   
\beqa   
T^4/\mbz_2:\qquad &&\Tr\g_{0,9}=\Tr\g_{0,5}=32,\cr   
&&\g_{\Om_0,9}=\g^T_{\Om_0,9}\ ,\qquad    
\g_{\Om_0,5}=-\g^T_{\Om_0,5},\cr   
&&\Tr\g_{1,9}=\Tr\g_{1,5}=0,\cr   
\cr   
T^4/\mbz_4:\qquad &&\Tr\g_{0,9}=\Tr\g_{0,5}=32,\cr   
&&\g_{\Om_0,9}=\g^T_{\Om_0,9}\ ,\qquad    
\g_{\Om_0,5}=-\g^T_{\Om_0,5},\cr   
&&\Tr\g_{1,9}=\Tr\g_{2,9}=\Tr\g_{1,5}=\Tr\g_{2,5}=0.   
\label{bzero}   
\eeqa   
The solutions are given by  
\beqa   
T^4/\mbz_2:\qquad &&\g_{1,9}=\g_{1,5}=\biggl(i I_{16},-i I_{16}\biggr),\cr   
\cr   
T^4/\mbz_4:\qquad &&\g_{1,9}=\g_{1,5}=\biggl(\b I_{8},\b^3 I_{8},   
\overline{\b} I_8,\overline{\b}^3 I_8\biggr),\qquad \b=e^{\pi i/4}.   
\label{bzerosolv}   
\eeqa     
where $I_n$ denotes an identity matrix of rank $n$.   
Furthermore, we can choose  $\g_{\Om_0,9}$ to be purely    
real and $\g_{\Om_0,5}$ purely imaginary. Eqs.~(\ref{bzero})   
then imply   
\beq   
\g_{\Om_0,9}=\pmatrix{0_{16}&I_{16}\cr   
I_{16}&0_{16}\cr},\qquad \g_{\Om_0,5}=\pmatrix{0_{16}&i I_{16}\cr   
-i I_{16}&0_{16}\cr}.   
\label{omegasol}   
\eeq   
The tadpole cancelation (\ref{bzero}) requires the introduction of 32
9-branes and 32 $5$-branes.  In principle, one can directly solve the
constraints (\ref{cons99})-(\ref{cons59}) on the $32\times 32$
hermitian Chan Paton matrices to determine the open string spectrum of
the model.  Alternatively, when the $\lambda$ matrices are written in
a Cartan-Weyl basis, the constraints on the Chan Paton matrices become
restrictions on the weight vectors. In this formalism \cite{ibanez},
the 99-sector gauge bosons correspond to both the Cartan generators
which trivially satisfy the $\lambda ^{(0)}$ constraint, together with
charged generators belonging to a subset of $SO(32)$ root vectors
selected by
\begin{equation} \rho^a \cdot V_{(99)}= 0 {\rm \, mod \,} \mbz,   
\label{v9p}   
\end{equation}   
where $V_{(99)}$ is the ``shift'' vector of the gauge twist $\g_{1,9}$   
\beq   
\gamma _{1,9}= e^{-2i\pi V_{(99)} \cdot H },   
\label{Vdef}   
\eeq   
and $\rho^a=\underline{(\pm1,\pm1,0,\cdots,0)},\ a=1,\cdots,480$   
are $SO(32)$ roots. (Here, underlining indicates that    
all permutations must be considered).   
In eq.~(\ref{Vdef}) $H$ is a vector of $SO(32)$ Cartan generators    
$\{H_I\},\ I=1,\cdots, 16$ and $V_{(99)}$ is determined from    
(\ref{bzerosolv}) to be    
\beqa   
T^4/\mbz_2:\qquad &&V_{(99)}={1\over 4}(1,1,1,1,1,1,1,1,1,1,1,1,1,1,1,1),\cr   
\cr   
T^4/\mbz_4:\qquad &&V_{(99)}={1\over 8}(1,1,1,1,1,1,1,1,3,3,3,3,3,3,3,3).   
\label{99shift}   
\eeqa   
Similarly, from the equation for $\lambda ^{(i)}$ in (\ref{cons99}) it   
follows that matter states correspond to charged generators with   
\begin{equation} \rho^a \cdot V_{(99)}= v_i {\rm \, mod \,} \mbz.   
\label{m9p}   
\end{equation}   
Since $\g_{\Omega_0,5}$ is antisymmetric,    
$\Omega$ projection in the 55-sector constraints    
gauge bosons to both the Cartan generators plus charged generators of    
$Sp(32)$ satisfying   
\beq   
\tilde{\rho}^a \cdot V_{(55)}= 0 {\rm \, mod \,} \mbz,   
\label{v5p}   
\eeq   
where $\tilde{\rho}^a$ include all the $SO(32)$ roots plus    
long roots $(\pm2,0,\cdots,0)$ and $V_{(55)}=V_{(99)}$.   
Matter fields in the 55-sector correspond to charged generators    
with    
\beqa    
{\rho}^a \cdot V_{(55)}&=& v_i {\rm \, mod \,} \mbz.   
\label{m5p}   
\eeqa   
Note that only a subset of short roots is used in determining    
the matter content of the 55-sector.    
This  is due to the     
extra minus sign in (\ref{cons55}) compare to the corresponding    
projection in (\ref{cons99}).    
Finally, the 95 sector is handled using an auxiliary $SO(64)\supset    
SO(32)_{(99)}\otimes SO(32)_{(55)}$ algebra.    
Since we have generators acting simultaneously on both 9-branes and   
5-branes,  only roots of the form   
\beq   
W _{(95)}= W_{(9)}\otimes W_{(5)}=   
({\underline {\pm 1, 0,  \cdots, 0}};{ \underline {\pm 1, 0, \cdots, 0}})   
\label{w95def}   
\eeq   
must be considered. Here the first (second) $16$ components transform
 under $SO(32)_{(99)}$ ($SO(32)_{(55)}$).  The shift in this sector is
 defined to be $V_{95}= V_{(99)}\otimes V_{(55)}$.  From
 (\ref{cons59}) we learn that massless states correspond to $W_{(95)}$
 roots satisfying
\begin{equation}   
W _{(95)} \cdot V _{(95)}=0 {\rm \, mod \,} \mbz.   
\label{95sh}   
\end{equation}   
The open string spectrum of the $T^4/\mbz_N$ orientifolds    
for $b=0$ is included in  Table \ref{6dtable1}.   
Our convention for the  $U(1)$ charges,   
as  explained in Appendix~\ref{lambdatraces}, differs from that  
in \cite{gp,gj9606}.     
   
Consider now the tadpoles of $T^4/\mbz_N$ orientifolds with non-zero
B-flux, $b\ne 0$.  Tadpoles with a volume factor $V_6 V_{T^4}$,
correspond to the untwisted R-R 10-form exchange in the closed string
tree channel.  All three surfaces KB, MS and C contribute to the
tadpole.  The KB contribution comes from the $k=0$ untwisted closed
string sector.  Since the parity projection $\Omega$ in this sector
requires all the windings to be zero, from (\ref{bmomentum}) we
conclude that the compact momentum sum does not change, and the
overall tadpole is the same as in the absence of the $B_{ij}$:
\beq   
\ck^{\rm 10-form}_u=(1-1)i\int_0^{\infty}\ {dt\over t^2}\ \biggl(   
{32\over N} {V_6 V_{T^4}\over {(4\pi^2\a')}^5}\biggr).   
\label{bklien}   
\eeq     
The C contribution comes from the $k=0$ 99-string    
sector. The partition function in this sector involves    
the summation over the 99-string  momentum modes. Upon   
Poisson resummation, which is necessary in order to extract    
the tadpole, the momentum sum in the open string loop channel is    
reinterpreted as a sum over the closed string    
winding modes in the  tree channel. Unlike the momentum    
states which remain unaffected by the $B_{ij}$ background,   
only  winding states with even winding numbers are allowed    
along the compact directions with $B_{ij}\ne 0$.    
Indeed, closed string winding states satisfy    
\beq   
P_{L}=-P_{R},   
\label{winding}   
\eeq   
which with the quantization (\ref{bmomentum}) and for a generic    
choice of metric on $T^4$ implies   
\beq   
n_i-b_{ij}m^j=0.   
\label{windingquant}   
\eeq     
Eq.~(\ref{windingquant}) requires $m^j\in 2\mbz$ for compact directions   
$j$ such that $b_{ij}\ne 0$. Even winding numbers in the    
closed tree channel further imply an effective reduction by half    
the basis momentum lattice vectors along these directions compare    
to the $b=0$ case \cite{Bij,w9712}.   
In the nontrivial  $B_{ij}$ background, the C tadpole is thus   
increased by $2^b$       
\beq   
\cc^{\rm 10-form}_u=i(1-1)\int_0^{\infty}\ {dt\over t^2}\ 
\biggl({2^b\over 16N}   
{V_6 V_{T^4}\over {(4\pi^2\a')}^5}\ {(\Tr\g_{0,9})}^2\biggr).   
\label{bc}   
\eeq     
Alternatively, the $2^b$ factor of the C tadpole can be interpreted    
as a sum over $2^b$ different sub-lattices within the $b=0$ case    
unit volume, shifted with respect to each other by half the    
generating vectors. A priori, the MS tadpole can have a    
different $\Omega$ projection $\g_{\Om_0,9}=\pm\g^T_{\Om_0,9}$,    
depending on a specific sub-lattice.     
In fact, the tadpole factorization requires   
that the MS contribution  increases by $2^{b/2}$, which    
is achieved by choosing the same $\Omega$ projection   
for $(2^{b-1}+2^{b/2-1})$ sub-lattices and the opposite projection    
for the other $(2^{b-1}-2^{b/2-1})$ sub-lattices   
\beq   
\cam^{\rm 10-form}_u=i(1-1)\int_0^{\infty}\ {dt\over t^2}\ \biggl(   
\mp{2^{b/2}\over N} {V_6 V_{T^4}\over {(4\pi^2\a')}^5}\ 
(\Tr\g_{0,9})\biggr),   
\label{bm}   
\eeq     
where a choice of the sign correlates with the choice of sign in $\Om$   
projection  for the larger set of momentum sub-lattices.    
Using the loop/tree channel moduli translation (\ref{tvsl}),   
the full tadpole of the untwisted  10-form  becomes   
\beqa   
[\ck+\cc+\cam]^{\rm 10-form}_u=i(1-1)\int_0^{\infty} d\ell    
{1\over 8N}{V_6 V_{T^4}\over {(4\pi^2\a')}^5}\    
\biggl[&&32^2\mp 2\cdot 2^{b/2}\cdot 32\ \Tr\g_{0,9}\cr  
&&+2^b\cdot    
{(\Tr\g_{0,9})}^2\biggr].   
\label{full10}   
\eeqa       
Tadpoles with a volume factor $V_6/V_{T_4}$ represent the    
untwisted R-R 6-form exchange in the closed string tree channel.   
The KB contribution comes from the $k=N/2$ untwisted closed    
string sector. In this sector the parity projection requires all    
the momenta to be zero and the NS-NS background     
allows only even windings along the compact directions with    
$b_{ij}\ne 0$. The net effect of the nontrivial background    
is thus a reduction of the corresponding $b=0$ tadpole   
by $2^b$   
\beq   
\ck^{\rm 6-form}_u=(1-1)i\int_0^{\infty}\ {dt\over t^2}\ \biggl(   
{32\over N\cdot 2^b}\ {V_6\over 4\pi^2\a' V_{T^4}}\biggr).   
\label{bklien6}   
\eeq     
Since $\ck^{\rm 6-form}_u$ tadpole measures the square of the total   
R-R charge of the sixteen orientifold 5-planes, its reduction    
compare to the zero NS-NS background tadpole implies that   
$2^{4-b}\cdot (2^{b-1}+2^{b/2-1})$ $O_5$-planes have the same charge    
as in the $b=0$ case while the other $2^{4-b}\cdot (2^{b-1}-2^{b/2-1})$ 
$O_5$-planes turn into $\tilde{O}_5$-planes which carry R-R charge 
opposite to that of the $O_5$-plane \cite{w9712}.    
The C contribution comes from the $k=0$ 55-string sector. Here,   
the partition function sums over the 55-string winding modes.    
The 55-string winding modes in the open string loop channel    
are reinterpreted upon the Poisson resummation as momentum    
modes in the closed string tree channel. Since the latter    
are insensitive to the NS-NS tensor background, the    
C tadpole is unchanged   
\beq   
\cc^{\rm 6-form}_u=i(1-1)\int_0^{\infty}\ {dt\over t^2}\ \biggl({1\over 16N}   
\ {V_6\over 4\pi^2\a' V_{T^4}}\ {(\Tr\g_{0,5})}^2\biggr).   
\label{bc6}   
\eeq     
The MS amplitude relevant to the 6-form exchange in the  tree channel is    
interpreted as a closed string exchange between  5-branes and sixteen    
orientifold 5-planes, and so is proportional to the total R-R charge   
of the orientifold 5-planes. It is suggested by (\ref{bklien6}) that    
the absolute value of the R-R charge of    
the orientifold 5-planes is reduced by $2^{b/2}$, so   
\beq   
\cam^{\rm 6-form}_u=i(1-1)\int_0^{\infty}\ {dt\over t^2}\ \biggl(   
\mp{1\over N\cdot 2^{b/2}}\  {V_6\over 4\pi^2\a' V_{T^4}}\ 
(\Tr\g_{0,5})\biggr).   
\label{bm6}   
\eeq     
where a choice of a sign correlates with the choice of a sign in 
$\Om$ projection in the 55-open string sector. 
The full tadpole of the untwisted  6-form in terms of    
the tree channel cylinder  modulus $\ell$ (\ref{tvsl}) becomes   
\beq   
[\ck+\cc+\cam]^{\rm 6-form}_u=i(1-1)\int_0^{\infty} d\ell    
{1\over 8N}\ {V_6\over 4\pi^2\a' V_{T^4}}\    
\biggl[{32^2\over 2^b}\mp {2\cdot 32\over 2^{b/2}}\ \Tr\g_{0,5}+    
{(\Tr\g_{0,5})}^2\biggr].   
\label{full6}   
\eeq       
The untwisted tadpole cancelation from (\ref{full10}) and (\ref{full6})    
requires    
\beqa   
\Tr\g_{0,9}&=&\Tr\g_{0,5}={32\over 2^{b/2}},   
\label{gammatraceun}   
\eeqa    
and the same $\Omega$ projection for both the larger set of the    
momentum sub-lattices in the 99-open string sector and the 55-open string 
sector as in the $b=0$ case:  
\begin{equation}  
\g_{\Om_0,9}=\g^T_{\Om_0,9}\ ,\qquad    
\g_{\Om_0,5}=-\g^T_{\Om_0,5}.   
\label{ompro}  
\end{equation}  
Since traces of $\g_{0,p}$ count the number of D-branes,   
(\ref{gammatraceun}) implies the rank reduction of a gauge    
group in the open string sector when the B-flux is turned on   
\cite{Bij,w9712,kst9803}.   
   
In addition to the untwisted R-R 6- and 10-form tadpoles ---   
which are canceled as discussed above,   
the orientifolds of interest have potential tadpoles        
corresponding to the twisted sector R-R 6-form exchange in    
the tree channel. These tadpoles  come from    
the cylinder vacuum amplitudes $C_{pq},\ p,q=\{5,9\}$, (\ref{camp})      
with $k\ne 0$ and so do not get contributions    
from momentum and winding modes on $T^4$.    
Since the nonzero NS-NS tensor background effects    
only the quantization of the momentum and winding modes on $T^4$, one    
would expect  the twisted R-R 6-form tadpole to be the same    
as in the $b=0$ case. As we will discuss in detail    
in Sec.~\ref{anomaly}, anomaly cancelations in six dimensions  
requires the $2^{b/2}$ multiplicity of massless    
states in the 59 open string sector. In Type IIB orientifolds,  
gauge degrees of freedom are carried by the Chan-Paton   
indices, so the multiplicity of massless states is extended to the    
multiplicity of Chan-Paton indices in the 59-sector and thus    
the  multiplicity of massive states in this sector as well.         
The latter implies that the cylinder vacuum amplitudes $\cc_{99}$,   
$\cc_{55}$ are unaffected by the B-flux while $\cc_{95}$ and    
$\cc_{59}$ get an extra factor of $2^{b/2}$. This raises a question:    
once the relative factor of $\cc_{99}$ and $\cc_{95}$ vacuum amplitudes   
is changed in the nonzero NS-NS background,    
how one maintains the factorization of the twisted R-R 6-form tadpoles?   
Recall that going to the tree channel interpretation in the    
KB vacuum amplitude we related (for $b=0$) the tree channel modulus $\ell$   
and the loop channel modulus $t_{\ck}$ through $\ell=1/(4t_{\ck})$.   
This KB tree channel ``cylinder'' has length $\ell$ and two crosscaps.   
The C and MS vacuum amplitudes of the open string sector in the tree    
channel picture were represented by an identical length cylinder and    
a cylinder with a crosscap. The fact that we used the same tree    
channel modulus for C and MS implicitly assumed the ability to relate    
KB, C and MS amplitudes to each other by adding/removing crosscaps.    
This is possible only for $\cc_{99}$ and $\cc_{55}$ since only they have    
the same type boundaries, which allows putting the crosscaps.    
The above arguments do not fix the relation between $\ell$    
and $t_{\cc_{59}}$. We claim that this relation must be fixed    
by demanding anomaly cancelation along with factorization    
of twisted R-R 6-form tadpoles. Thus we require    
\beq   
\ell={2^{b-1}\over t_{\cc_{59}}}.   
\label{tc59}   
\eeq        
The twisted R-R 6-form tadpoles vanish provided   
\beqa   
T^4/\mbz_2:\qquad    
&&\Tr\g_{1,9}=\Tr\g_{1,5}=0,\cr   
T^4/\mbz_4:\qquad &&\Tr\g_{1,9}=\Tr\g_{2,9}=\Tr\g_{1,5}=\Tr\g_{2,5}=0.   
\label{bnonzero}   
\eeqa   
Eqs.~(\ref{gammatraceun}) and (\ref{bnonzero}) are solved by    
\beqa   
T^4/\mbz_2:\qquad &&\g_{1,9}=\g_{1,5}=\biggl(i I_{2^{4-b}},-i    
I_{2^{4-b}}\biggr),\cr   
\cr   
T^4/\mbz_4:\qquad &&\g_{1,9}=\g_{1,5}=\biggl(\b I_{2^{3-b}},\b^3 I_{2^{3-b}}, 
\overline{\b} I_{2^{3-b}},\overline{\b}^3 I_{2^{3-b}}\biggr),\qquad    
\b=e^{\pi i/4}.   
\label{bnonzerosolv}   
\eeqa     
The ``shift'' vectors corresponding to the above gauge twists are   
\beqa    
T^4/\mbz_2:\qquad &&V_{(99)}={1\over 4}(\ \overbrace{1,\cdots,1}^{2^{4-b}\   
{\rm times}}\ ),\cr   
\cr   
T^4/\mbz_4:\qquad &&V_{(99)}={1\over 8}(\ \underbrace{1,\cdots,1}_{2^{3-b}\   
{\rm times}},\ \underbrace{3,\cdots,3}_{2^{3-b}\ {\rm times}}\ ).   
\label{shiftz4}   
\eeqa   
with  $V_{(55)}=V_{(99)}$ and $V_{(95)}=V_{(99)}\otimes V_{(55)}$.   
As for the $b=0$ case,  we can choose  $\g_{\Om_0,9}$ purely    
real and $\g_{\Om_0,5}$ to be purely imaginary. Eqs.~(\ref{ompro})   
then imply   
\beq   
\g_{\Om_0,9}=\pmatrix{0_{2^{4-b}}&I_{2^{4-b}}\cr   
I_{2^{4-b}}&0_{2^{4-b}}\cr},\qquad    
\g_{\Om_0,5}=\pmatrix{0_{2^{4-b}}&i I_{2^{4-b}}\cr   
-i I_{2^{4-b}}&0_{2^{4-b}}\cr}.   
\label{omegasolun}   
\eeq   
Given (\ref{bnonzerosolv}) and (\ref{omegasolun}),   
the massless open string states can be determined as in the    
$b=0$ case. The open string spectra are  
given in Table \ref{6dtable1}.

\subsection{Closed String Sector}   
\label{closedsector}    
In this section we discuss the closed string sector of Type IIB   
orientifolds on $T^4/\mbz_N$, ($N=2,4$), in the presence of B-flux.    
   
First consider the $\mbz_2$ orientifold models.    
The untwisted sector of the models is constructed by keeping states    
of Type IIB orbifold on $T^4/\mbz_2$ invariant under the world sheet    
parity $\Om$. This gives rise to a $\cn=1$ supergravity multiplet in    
six dimensions, accompanied by one tensor multiplet and   
four hypermultiplets.   
   
The twisted sectors will produce additional multiplets. The bosonic    
content of a hypermultiplet is four scalars transforming as    
$4{\bf (1,1)}$ under the six dimensional little group    
$SO(4)\approx SU(2)\otimes SU(2)$, while that of a tensor multiplet    
as ${\bf (1,1)}\oplus {\bf (1,3)}$. There are overall 16 fixed points    
in the $\mbz_2$ orbifold with an  orientifold 5-plane sitting    
at each fixed point. As we explained in Sec.~\ref{opensector},   
out of 16 orientifold 5-planes, each one of $2^{4-b}\cdot (2^{b-1}+2^{b/2-1})$
$O_5$-planes carries $(-2)$ units of R-R charge while the    
other $2^{4-b}\cdot (2^{b-1}-2^{b/2-1})$ $\tilde{O}_5$-planes have    
opposite R-R charge. This implies that precisely    
$2^{4-b}\cdot (2^{b-1}+2^{b/2-1})$ $\mbz_2$ fixed points    
are even under the $\Om$ projection while the other    
$2^{4-b}\cdot (2^{b-1}-2^{b/2-1})$   
$\mbz_2$ fixed points pick up a minus sign under the $\Om$ projection.  
At each fixed point of the twisted sector, the NS and R sector    
massless states transform as   
\begin{equation}   
\begin{array}{lc}   
{\mbox {Sector}} \quad \quad & SU(2) \otimes SU(2) ~ {\mbox {rep.}} \\   
{\mbox {NS}}     & 2 ({\bf 1},{\bf 1})~, \\   
{\mbox {R}}      & ({\bf 1},{\bf 2})~. \\   
\end{array}   
\end{equation}   
The twisted sector spectrum is obtained by taking products of states   
from the left- and right-moving sectors. The orientifold projection   
$\Omega$ keeps symmetric (antisymmetric) combinations in the NS-NS   
sector and antisymmetric (symmetric) combinations in the R-R sector   
for each $\Om$-even ($\Om$-odd) $\mbz_2$ fixed point. This gives the   
bosonic content of a hypermultiplet for each $\Om$-even $\mbz_2$ fixed   
point and the bosonic content of a tensor multiplet for each $\Om$-odd   
$\mbz_2$ fixed point. Overall, the twisted sectors of $\mbz_2$ orientifold   
contribute $2^{4-b}\cdot (2^{b-1}+2^{b/2-1})$ hypermultiplets    
and $2^{4-b}\cdot (2^{b-1}-2^{b/2-1})$ extra tensor multiplet  
(see Table \ref{6dtable1}).   
   
The untwisted sector of $T^4/\mbz_4$ orientifold contains $\cn=1$   
supergravity multiplet in six dimensions, accompanied by one tensor   
multiplet and two hypermultiplets.    
     
The four $\mbz_4$ invariant fixed points of $T^4/\mbz_4$ orientifold   
give four hypermultiplets and four tensor multiplet.    
They are also $\mbz_2$ fixed points and so supply an additional   
4 hypermultiplets. The other 12 $\mbz_2$ fixed points form    
6 $\mbz_4$ invariant pairs. Out of 12 $\mbz_2$ fixed points    
which are not $\mbz_4$ invariant, $2^{4-b}\cdot (2^{b-1}-2^{b/2-1})$    
fixed points are odd under the $\Om$ projection and the    
other $12-2^{4-b}\cdot (2^{b-1}-2^{b/2-1})$ are even.    
Thus they contribute $2^{3-b}\cdot (2^{b-1}-2^{b/2-1})$   
extra tensor multiplets and $6-2^{3-b}\cdot (2^{b-1}-2^{b/2-1})$   
hypermultiplets. Overall, twisted sectors of $\mbz_4$ orientifold    
contribute $4+2^{3-b}\cdot (2^{b-1}-2^{b/2-1})$ extra tensor    
multiplets and  $14-2^{3-b}\cdot (2^{b-1}-2^{b/2-1})$   
hypermultiplets (see Table \ref{6dtable1}).

\section{Anomaly Cancelations}   
\label{anomaly}   
An important check on the consistency of the $\cn=1$ six dimensional    
vacua constructed in the previous sections is provided by the    
cancelation of all gauge, gravitational, and mixed    
anomalies. Cancelation of tadpoles in Type IIB orientifolds    
on $T^4/\mbz_N$ ($N=2,4$) discussed in Sec.~\ref{opensector}   
guarantees 1-loop consistency of the equations of motion   
for the untwisted R-R 6- and 10-form as well as the twisted R-R   
6-form. Recall that the tadpoles were extracted from the    
KB, MS and C vacuum amplitudes. Each of these amplitudes    
involves a summation over four spin structures ({\sl i.e.},    
summation over $\{\a,\b\}=\{0,\oh\}$ in Appendix~\ref{1loopb0}).    
Actually, the R-R spin structure,    
$\{\a,\b\}=\{\oh,\oh\}$ in Appendix~\ref{1loopb0}, does not    
contribute to the 1-loop vacuum amplitudes because of the    
fermionic zero modes. On the other hand, it will contribute    
to the  2-loop vacuum amplitudes, as well as to various scattering    
processes. In fact, being odd under the spacetime parity,    
it is the only spin structure (at 1-loop level) that would generate   
potential chiral anomalies in the low-energy effective field    
theory.     
In this sense, the cancelation of six dimensional chiral    
anomalies follows from the absence of tadpoles coming from the    
R-R spin structure. In this section we show that $T^4/\mbz_N$ ($N=2,4$)    
Type IIB orientifolds with B-flux are free from chiral anomalies.    
First, we explain anomaly cancelation in the framework of    
the six dimensional effective field theory. Then we discuss  
how the Green-Schwarz terms can be computed from the corresponding  
non-planar diagrams in string theory.  
In the latter    
approach we confirm the necessity of $2^{b/2}$ multiplicity of    
59-cylinder diagrams. Some useful facts on six dimensional chiral anomalies    
are given in Appendix~\ref{anomalyreview}.     
   
\subsection{$\mbz_2$ Orientifolds}   
   
We start with the $T^4/\mbz_2$ orientifold models. From the    
spectrum of the models given in Table \ref{6dtable1}   
it is straightforward to check using the  
results of Appendix~\ref{anomalyreview}    
that the  
leading anomalies, {\em i.e.}, the $\tr R^4$ and $\tr F^4$ terms 
vanish.  
To see this explicitly, 
notice that the leading term in the  
pure gravitational anomalies is proportional to 
\begin{equation}\label{R4} 
(n_H - n_V + 29 n_T -273) ~ \tr R^4, 
\end{equation} 
where $n_H$ is the number of hypermultiplets from both open 
and closed string sectors. 
From Section \ref{closedsector}, $n_T-1=8-8/2^{b/2}$ and the 
number of closed string hypermultiplets is $n_H^c=12+8/2^{b/2}$. 
The gauge group is $U(N) \times U(N)$ where 
$N=16/2^{b/2}$, and hence $n_V=2 N^2$. The number of open string 
hypermultiplets is $n_H^o = 2 N(N-1) + 2^{b/2} N^2$. 
It is easy to see that the above term (\ref{R4}) vanishes. 
Notice that it is crucial for the cancelation of    
the $\tr R^4$ gravitational anomaly that the multiplicity    
of 59-open string matter states is $2^{b/2}$.  
In addition, the    
hypermultiplet representations plus the $2^{b/2}$ multiplicity    
in the 59-sector ensure that the $\tr F^4$ term is absent. 
This can be seen from (\ref{tracesrelations}) that the 
leading pure gauge non-abelian anomalies for each $SU(N)$ gauge group 
is proportional to 
\begin{equation}   
(2 N - 2 (N-8) - 2^{b/2} N)~ \tr F^4 
\end{equation}  
which vanishes since the multiplicity of $59$ states is $2^{b/2}$ 
which is equal to $16/N$. 
The remaining gravitational $+$ gauge anomaly may be written    
in the form    
\beq   
i{(2\pi)}^3 I=-{1\over 16\cdot 2^{b/2}}\ X_4^{(9)}\wedge X_4^{(5)}   
+X_2\wedge X_6^{(9)}-\tilde{X}_2\wedge X_6^{(5)},   
\label{z2anomaly}   
\eeq   
where    
\beqa   
X_4^{(a)}&=&R^2-2\cdot 2^{b/2}\ (F_a^2+f_a^2),\cr   
X_6^{(a)}&=&{1\over 6}   
\bigg[R^2 f_a-2^{b/2}f_a^3-3\cdot 2^{b/2} F_a^2 f_a-4\cdot 2^{b/4} 
F_a^3\bigg],   
\label{46z2}   
\eeqa   
and    
\beqa   
X_2&=&{4\over 2^{b/2}}f_9-f_5,\cr   
\tilde{X}_2&=&{4\over 2^{b/2}}f_5-f_9.   
\label{2z2}   
\eeqa   
The subscript $a=5,9$ refers to the gauge group factors in    
the 55- and  99-open string sectors\footnote{The anomalies   
of the $b=0$ models were discussed previously   
in \cite{6authors}. The difference   
between the corresponding anomaly polynomials   
of (\ref{2z2}) and the one given in \cite{6authors}  
is due to a different convention of the $U(1)$ charges.   
See Table \ref{6dtable1} for our convention.}.    
The anomaly polynomial    
(\ref{z2anomaly}) contains terms that involve only nonabelian    
field strength as well as terms that also involve abelian field strength.    
The former can be canceled by exchange of a two-form,    
which involves Green-Schwarz interactions of the form   
\beqa   
\Gamma_{c.t.}={i\over 16\  2^{b/2}\ {(2\pi)}^3}\int B_2\wedge X_4^{(5)}.   
\label{counter1}   
\eeqa   
The $B_2$ field is the 2-form with gauge invariant    
field strength $H=d B_2-X_3^{(9)}$ where $X_4^{(9)}=d X_3^{(9)}$.       
The counterterm (\ref{counter1}) arises from the tree-level diagram    
which involves coupling of the $B_2$ field and a pair    
of gauge bosons. Such couplings can be computed    
from the disk diagram in string theory.       
When $b\ne 0$, a disk diagram involving two gauge bosons and a two-form    
potential of the twisted sector tensor multiplet is proportional    
to $\tr(\g_{1,p}{(\lambda_p)}^2)$, where $p=5,9$ refers to the    
55- or 99-open string sector and $\lambda_p$ denotes the Chan-Paton   
generator of the corresponding gauge boson. Using the results in    
Appendix~\ref{lambdatraces}, this trace is zero.       
Thus we identify the 2-form $B_2$  as a combination of the two-form   
potential of the gravity multiplet with a self-dual field strength    
and the two-form potential with an anti-self-dual field strength    
of the untwisted sector tensor multiplet. The remaining terms    
of the anomaly polynomial (\ref{z2anomaly}) can be canceled by    
\begin{equation}\label{counterBL}\Gamma'_{c.t.}=-{i\over {(2\pi)}^3}   
\int B_0^{(9)}\;    
X_6^{(9)}+{i\over {(2\pi)}^3}\int B_0^{(5)}\; X_6^{(5)},   
\end{equation}   
if we assign the anomalous transformation laws:   
\begin{eqnarray}\label{anomshift}   
B_0^{(9)}\rightarrow B_0^{(9)}+ {4\over 2^{b/2}}\epsilon_9-\epsilon_5,     
\\B_0^{(5)}\rightarrow B_0^{(5)}+ {4\over 2^{b/2}}\epsilon_5-\epsilon_9,     
\nonumber   
\end{eqnarray}   
under the $U(1)_a$ gauge transformations $A_a\rightarrow A_a
+d\epsilon_a$.  As in (\ref{counter1}), the counterterm
(\ref{counterBL}) arises from a tree-level diagram. This diagram
involves the coupling of the $B_0^{(a)}$ scalars with a single
$U(1)_a$ gauge boson. From the disk diagram, the coupling of the
untwisted sector R-R scalars to $U(1)_a$ gauge boson is proportional
to $\tr (\lambda_a)$ which vanishes, see Appendix~\ref{lambdatraces}.
On the other hand, the corresponding coupling of the twisted sector
R-R scalars is proportional to $\tr \g_{1,a}\lambda_a\ne 0$.
   
Consider first the $b=0$ case.  The number of $\mbz_2$ twisted scalars
$\phi_I$ is $16$. All of them couple with the same relative strength
to the 99-sector $U(1)$, because $\tr(\g_{1,9}\lambda_9)$ (which
measures the coupling strength) is the same for all fixed
points. Since all D5-branes sit at the origin, only the twisted scalar
at the origin (which we denote by $\phi_1$) will couple. This can be
seen explicitly from the tadpole solution: the coupling of the scalar
at a fixed point $I$ to a 55-sector gauge boson is proportional to
$\tr(\g_{1,5,I}\lambda_5)$, which is zero unless $I=1$ is the origin,
where all the D5-branes are located.  Therefore,
\beqa   
B_0^{(9)}&=&\a \sum_{I=1}^{16}\phi_I,\cr   
B_0^{(5)}&=&\b\ \phi_1,   
\label{b0z2}   
\eeqa   
for some coefficients $\a$ and $\b$ to be determined.   
Since $\phi_1$ couples both to 55-sector and 99-sector    
gauge fields, it transforms under $U(1)$ gauge transformations    
as   
\beq   
\phi_1\to \phi_1+a_1\ep_9+a_2\ep_5,   
\label{f1z2}   
\eeq    
for some coefficients $a_1$ and $a_2$ to be determined.   
The other scalars $\phi_I,\ I=2,\cdots 16$, couple only to    
the 99-sector, but all in a same way, so   
\beq   
\phi_I\to \phi_I+b_1\ep_9.   
\label{fiz2}   
\eeq    
Requiring that (\ref{b0z2})-(\ref{fiz2}) generate      
(\ref{anomshift}) uniquely fixes all the coefficients:   
$\a, \b, a_1, a_2, b_1$  up to    
field normalization. We have  
\beqa   
B_0^{(9)}&=&\a \sum_{I=1}^{16}\phi_I,\cr   
B_0^{(5)}&=&-4\a\ \phi_1,  
\eeqa  
where  
\beqa  
\phi_1&\to& \phi_1+(\ep_9-4\ep_5)/(4\a),\cr   
\phi_I&\to& \phi_I+\ep_9/(4\a),\qquad I=2,\cdots,16.   
\label{finalz2b0}   
\eeqa   
Since the $\phi_I$'s  are in  linear multiplets, the choice of normalization   
$\a$ is irrelevant. The choice $\a=1/4$ reproduces the result of   
\cite{6authors}. One can always attain the canonical normalization    
of $B_0^{(9)}$ and $B_0^{(5)}$ with an appropriate   
change of the coupling constants in the counterterm (\ref{counterBL}).    
This is also generic to other    
orientifold models: a linear combination of twisted scalars   
that couple to the abelian gauge fields is unique, up to field   
normalization.   
   
In the $b=2$ case,  the number of $\mbz_2$ twisted scalars is $12$.    
As in the previous case, one scalar is singled out as living    
at the fixed point where we put all D5-branes.    
Using the same logic as before, we find a unique solution   
\beqa   
B_0^{(9)}&=&\a\ \sum_{I=1}^{12}\phi_I,\cr   
B_0^{(5)}&=&-2\a\ \phi_1,  
\eeqa  
where  
\beqa   
\phi_1&\to& \phi_1+(\ep_9-2\ep_5)/(2\a),\cr   
\phi_I&\to& \phi_I+3 \ep_9/(22 \a),\qquad I=2,\cdots,12.   
\label{finalz2b2}   
\eeqa   
   
In the $b=4$ case, the number of $\mbz_2$ twisted scalars   
is $10$. Therefore, we find   
\beqa   
B_0^{(9)}&=&\a\ \sum_{I=1}^{10}\phi_I,\cr   
B_0^{(5)}&=&-\a\ \phi_1,  
\eeqa  
where  
\beqa   
\phi_1&\to& \phi_1+(\ep_9-\ep_5)/\a,\cr      
\phi_I&\to& \phi_I,\qquad I=2,\cdots,10.   
\label{finalz2b4}   
\eeqa   
   
\subsection{$\mbz_4$ Orientifolds}   
   
We now consider $T^4/\mbz_4$ orientifold models. From   
the open string spectrum of the models given in Table \ref{6dtable1} 
and the closed string spectrum in Section \ref{closedsector},    
one can easily see that 
the $\tr R^4$ and all $\tr F^4$ terms vanish in the 8-form anomaly     
polynomial. The remaining gravitational $+$ gauge anomaly may be    
written in the form   
 \beqa   
i{(2\pi)}^3 I=&&-{1\over 32\cdot 2^{b/2}} X_4^{(9)}\wedge X_4^{(5)}   
-{2^{b/2}\over 8}\tilde{X}_4^{(9)}\wedge\tilde{X}_4^{(5)}+   
{1\over 8}\tilde{X}_4^{(5)}\wedge\tilde{X}_4^{(5)}+   
{1\over 8}\tilde{X}_4^{(9)}\wedge\tilde{X}_4^{(9)}\cr   
\cr   
&&+X_2^{(1,9)}\wedge X_6^{(1,9)}+X_2^{(3,9)}\wedge X_6^{(3,9)}   
+X_2^{(1,5)}\wedge X_6^{(1,5)}+X_2^{(3,5)}\wedge X_6^{(3,5)},   
\label{anomz4}   
\eeqa   
where    
\beqa   
X_4^{(a)}&=&R^2-2\cdot 2^{b/2} (F_{1,a}^2+F_{3,a}^2+f_{1,a}^2+f_{3,a}^2),\cr   
\cr   
\tilde{X}_4^{(a)}&=&F_{1,a}^2+f_{1,a}^2-F_{3,a}^2-f_{3,a}^2,\cr   
\cr   
X_6^{(\a,a)}&=&{1\over 12\cdot 2^{b/2}}R^2 f_{\a,a}-{1\over 6}f_{\a,a}^3   
-{1\over 2}F_{\a,a}^2 f_{\a,a}-{2^{(2-b)/4}\over 3} F_{\a,a}^3,   
\label{46z4}   
\eeqa   
and    
\beqa   
X_2^{(1,9)}&=&3 f_{1,9}-f_{3,9}-2^{b/2} f_{1,5},\cr   
X_2^{(3,9)}&=&3 f_{3,9}-f_{1,9}-2^{b/2} f_{3,5},\cr   
X_2^{(1,5)}&=&3 f_{1,5}-f_{3,5}-2^{b/2} f_{1,9},\cr   
X_2^{(3,5)}&=&3 f_{3,5}-f_{1,5}-2^{b/2} f_{3,9}.   
\label{2z4}   
\eeqa   
In the $Z_4$ orientifolds there are two nonabelian    
gauge factors in the 99-sector and 55-sector. One comes    
from the shift-vector $1/8$ components  and the other    
come from the shift-vector $3/8$ components, (c.f. (\ref{shiftz4})).    
In (\ref{anomz4}), we  use this    
to label different gauge factors. For example,   
the 99-sector nonabelian field strengths will    
be denoted by $F_{1,9}$ and $F_{3,9}$. The nonabelian    
factors in the 55-sector are denoted by   
$F_{1,5}$ and $F_{3,5}$.   
We use similar notation $f_{\a,a}$ for    
the abelian factors.   
The terms of the anomaly polynomial (\ref{anomz4})    
that involve only nonabelian field strengths    
can be canceled by exchange of three two-forms    
with Green-Schwarz interactions   
\beq   
\Gamma_{c.t.}={i\over 32\  2^{b/2}\ {(2\pi)}^3}\int    
B_2^{(1)}\wedge X_4^{(5)}+{i\ 2^{b/2}\over 8\   {(2\pi)}^3}\int    
B_2^{(2)}\wedge \tilde{X}_4^{(5)}-{i \over 8\   {(2\pi)}^3}\int    
B_2^{(3)}\wedge \tilde{X}_4^{(9)},   
\label{z42form}   
\eeq   
where $B_2^{(i)}$ 's have gauge invariant field strengths   
\beqa   
H^{(1)}&=&d B_2^{(1)} - X_3^{(9)},\qquad X_4^{(9)}=d X_3^{(9)},\cr   
H^{(2)}&=&d B_2^{(2)} - X_3^{(9,5)},\qquad    
\tilde{X}_4^{(9)}-{1\over 2^{b/2}}\tilde{X}_4^{(5)}=d X_3^{(9,5)},\cr   
H^{(3)}&=&d B_2^{(3)} - \tilde{X}_3^{(9)},\qquad    
\tilde{X}_4^{(9)}=d \tilde{X}_3^{(9)}.   
\label{2fromtransform}   
\eeqa   
The counterterm (\ref{z42form}) arises from a tree-level diagram    
which involves coupling of $B_2^{(i)}$ fields and a pair of    
gauge bosons. Recall that these models have a single    
tensor multiplet from the untwisted sector, $4$ tensor multiplets    
from the $\mbz_4$ twisted sector and    
$2^{3-b}\cdot(2^{b-1}-2^{b/2-1})$ extra tensor multiplets coming from    
the $\mbz_2$ twisted sector. By the same arguments as for the    
$T^4/\mbz_2$ orientifolds, the $\mbz_2$ twisted sector tensor multiplets    
do not couple to a pair of gauge bosons. On the contrary,   
the coupling of the $\mbz_4$ twisted sector tensor multiplets    
from the disk diagram is  nonzero. With all the D5-branes at the    
origin, only the $\mbz_4$ twisted tensor multiplet at the origin    
will couple to the 55-sector gauge bosons.    
Thus we expect that $B_2^{(1)}$ and $B_2^{(3)}$    
are combinations of the untwisted sector two-form potentials    
of the gravity and the tensor multiplets plus two-form    
potentials of all four $\mbz_4$ twisted sector tensor multiplets.   
$B_2^{(2)}$ must contain the untwisted sector two-form potentials   
and a  two-form potential from the $\mbz_4$ twisted sector tensor   
multiplet at the origin. The remaining abelian terms of the anomaly    
polynomial (\ref{anomz4}) may be canceled by   
additional counterterms    
\begin{eqnarray}\label{counterBLz4}\Gamma'_{c.t.}=&&   
-{i\over {(2\pi)}^3} \int B_0^{(1,9)}\;   
X_6^{(1,9)} - {i\over {(2\pi)}^3}\int B_0^{(3,9)}\;   
X_6^{(3,9)} - {i\over {(2\pi)}^3}\int B_0^{(1,5)}\; X_6^{(1,5)}\cr   
&&-{i\over {(2\pi)}^3} \int B_0^{(3,5)}\; X_6^{(3,5)},   
\end{eqnarray}   
if we assign the anomalous transformation laws:   
\begin{eqnarray}\label{anomshiftz4}   
B_0^{(1,9)}\rightarrow B_0^{(1,9)}+ 3\epsilon_{1,9}-\epsilon_{3,9}   
-2^{b/2}\ep_{1,5},   
\\B_0^{(3,9)}\rightarrow B_0^{(3,9)}+ 3\epsilon_{3,9}-\epsilon_{1,9}   
-2^{b/2}\ep_{3,5},   
\nonumber   
\\B_0^{(1,5)}\rightarrow B_0^{(1,5)}+ 3\epsilon_{1,5}-\epsilon_{3,5}   
-2^{b/2}\ep_{1,9},   
\nonumber   
\\B_0^{(3,5)}\rightarrow B_0^{(3,5)}+ 3\epsilon_{3,5}-\epsilon_{1,5}   
-2^{b/2}\ep_{3,9},   
\nonumber   
\end{eqnarray}   
under the $U(1)$ gauge transformations $A_{\a,a}\rightarrow A_{\a,a}
+d\epsilon_{\a,a}$.  

Notice that in the presence of D-branes, the total twist of a
non-vanishing
$N$-point function does not have to be zero. Therefore, 
the anomalous $U(1)$ gauge bosons 
can couple to closed string states from
different twisted sectors (in this case ${\mathbb Z}_2$ and
${\mathbb Z}_4$ twisted sectors). As a result, there are
mixings in the kinetic terms of the ${\mathbb Z}_2$ and
${\mathbb Z}_4$ twisted sector states even when $g_s =0$.
We now proceed to identify the scalars
$B_0^{(\a,a)}$ in terms of the twisted R-R states.
   
We start with the $b=0$ case.  There are in total 14 twisted scalars:
4 scalars (we denote them by $\phi_I^{(4)},\ I=1,\cdots 4$) coming
from the $\mbz_4$ twisted sector, the other 10 scalars (we denote them
$\phi_J^{(2)},\ J=1,\cdots 10$) coming from the $\mbz_2$ twisted
sector. Out of 10 $\mbz_2$ twisted fields, 4 come from the $\mbz_2$
fixed points which are also $\mbz_4$ fixed points, and the other 6
come from 6 pairs of the $\mbz_2$ fixed points which are $\mbz_4$
invariant.  Let us analyze the coupling of these R-R twisted fields to
the various $U(1)$ gauge bosons.  All twisted fields couple to the
99-sector vector bosons. These couplings can be computed from the disk
diagram: for the $\mbz_4$ twisted fields the coupling is proportional
to $\tr(\g_{1,9}\lambda_{\a,9})$ while for the $\mbz_2$ twisted fields
the coupling is proportional to $\tr(\g_{2,9}\lambda_{\a,9})$. Since
\beqa   
{\tr(\g_{1,9}\lambda_{1,9})\over \tr(\g_{1,9}\lambda_{3,9})}&=&1,\cr   
{\tr(\g_{2,9}\lambda_{1,9})\over \tr(\g_{2,9}\lambda_{3,9})}&=&-1,   
\label{t1}   
\eeqa   
the coupling of $\phi_I^{(4)}$ to the $A_{3,9}$ abelian    
gauge boson is  identical to that of the $A_{1,9}$,   
while the coupling of $\phi_J^{(2)}$ to $A_{3,9}$   
is opposite to its coupling to $A_{1,9}$.   
As in the case of $\mbz_2$ orientifold, only    
twisted fields that live at the fixed point where the    
D5-branes are located will couple to the 55-sector gauge bosons.   
We call these fields $\phi_1^{(4)}$ and $\phi_1^{(2)}$.   
Their coupling to different $U(1)$ can be deduced as discussed   
above.    
Collecting the information about the coupling,   
we can easily generalize (\ref{b0z2}) to:   
\beqa   
B_0^{(1,9)}&=&\a\sum_{I=1}^4 \phi_I^{(4)}+   
\b \sum_{J=1}^{10}\phi_J^{(2)},\cr   
\cr   
B_0^{(3,9)}&=&\a\sum_{I=1}^4 \phi_I^{(4)}-   
\b \sum_{J=1}^{10}\phi_J^{(2)},\cr   
\cr   
B_0^{(1,5)}&=&\g\  \phi_1^{(4)}+   
\d\ \phi_1^{(2)},\cr   
\cr   
B_0^{(3,5)}&=&\g\  \phi_1^{(4)}-   
\d\  \phi_1^{(2)},   
\label{b0z4}   
\eeqa   
where $\a,\b,\g,\d$  are coefficients to be determined.   
Using the same logic as for the $\mbz_2$ orientifolds,   
we assign the following gauge transformations to the    
twisted fields   
\beqa   
&&\phi_1^{(4)}\to \phi_1^{(4)}+a_1\ep_{1,9}+a_2\ep_{3,9}+a_3\ep_{1,5}+   
a_4\ep_{3,5},\cr   
&&\phi_I^{(4)}\to \phi_I^{(4)}+b_1\ep_{1,9}+b_2\ep_{3,9},   
\qquad\qquad\qquad\qquad    
I=2,\cdots 4,\cr   
&&\phi_1^{(2)}\to \phi_1^{(2)}+c_1\ep_{1,9}+c_2\ep_{3,9}+c_3\ep_{1,5}+   
c_4\ep_{3,5},\cr   
&&\phi_J^{(2)}\to \phi_J^{(2)}+d_1\ep_{1,9}+d_2\ep_{3,9},\qquad\qquad   
\qquad\qquad   
J=2,\cdots 10.   
\label{trans}   
\eeqa   
As we already mentioned, requiring that (\ref{b0z4})-(\ref{trans})   
satisfy (\ref{anomshiftz4}), fixes uniquely all coefficients    
up to the field normalization.  (The    
system is overconstraint and so generically    
would not have solution at all.) We end up with   
\beqa   
B_0^{(1,9)}&=&\a \sum_{I=1}^4 \phi_I^{(4)}+   
\b \sum_{J=1}^{10}\phi_J^{(2)},\cr   
\cr   
B_0^{(3,9)}&=&\a \sum_{I=1}^4 \phi_I^{(4)}-   
\b \sum_{J=1}^{10}\phi_J^{(2)},\cr   
\cr   
B_0^{(1,5)}&=&-2\a\ \phi_1^{(4)}-4\b\ \phi_1^{(2)},\cr   
\cr   
B_0^{(3,5)}&=&-2\a\ \phi_1^{(4)}+4\b\ \phi_1^{(2)},   
\eeqa  
where  
\beqa  
\phi_1^{(4)}&\to& \phi_1^{(4)}+(\ep_{1,9}+\ep_{3,9})/(4 \a)-(\ep_{1,5}+   
\ep_{3,5})/(2 \a),\cr   
\phi_I^{(4)}&\to& \phi_I^{(4)}+(\ep_{1,9}+\ep_{3,9})/(4 \a),   
\qquad   
I=2,\cdots 4,\cr   
\phi_1^{(2)}&\to& \phi_1^{(2)}+(\ep_{1,9}-\ep_{3,9})/(8 \b)-(\ep_{1,5}-   
\ep_{3,5})/(2 \b),\cr   
\phi_J^{(2)}&\to& \phi_J^{(2)}+5(\ep_{1,9}-\ep_{3,9})/(24 \b),\qquad   
J=2,\cdots 10.   
\label{finalz4b0}   
\eeqa   
The appearance of two normalization constants in the final    
expression comes from an independent choice of normalizations    
for the $\mbz_4$ and $\mbz_2$ twisted fields.

Turning on rank two NS-NS two form field $B_{ij}$ converts two    
of the $\mbz_2$ twisted hypermultiplets into tensor multiples, so the    
total number of $\mbz_2$ twisted scalars becomes $8$.    
Repeating identical analysis as in the $b=0$ case,    
we find a unique solution (up to field normalization)   
\beqa   
B_0^{(1,9)}&=&\a \sum_{I=1}^4 \phi_I^{(4)}+   
\b \sum_{J=1}^{8}\phi_J^{(2)},\cr   
\cr   
B_0^{(3,9)}&=&\a \sum_{I=1}^4 \phi_I^{(4)}-   
\b \sum_{J=1}^{8}\phi_J^{(2)},\cr   
\cr   
B_0^{(1,5)}&=&-\a\ \phi_1^{(4)}-2\b\ \phi_1^{(2)},\cr   
\cr   
B_0^{(3,5)}&=&-\a\ \phi_1^{(4)}+2\b\ \phi_1^{(2)},   
\eeqa  
where  
\beqa  
\phi_1^{(4)}&\to& \phi_1^{(4)}+(\ep_{1,9}+\ep_{3,9})/\a-(\ep_{1,5}+   
\ep_{3,5})/\a,\cr   
\phi_I^{(4)}&\to& \phi_I^{(4)},   
\qquad   
I=2,\cdots 4,\cr   
\phi_1^{(2)}&\to& \phi_1^{(2)}+(\ep_{1,9}-\ep_{3,9})/(2 \b)-(\ep_{1,5}-   
\ep_{3,5})/\b,\cr   
\phi_J^{(2)}&\to& \phi_J^{(2)}+3(\ep_{1,9}-\ep_{3,9})/(14 \b),\qquad   
J=2,\cdots 8.   
\label{finalz4b2}   
\eeqa   
   
Finally, the rank four B-flux converts three   
of the $\mbz_2$ twisted hypermultiplets into tensor multiples, so the   
total number of $\mbz_2$ twisted scalars becomes $7$.   
In this case we find   
\beqa   
B_0^{(1,9)}&=&\a \sum_{I=1}^4 \phi_I^{(4)}+   
\b \sum_{J=1}^{7}\phi_J^{(2)},\cr   
\cr   
B_0^{(3,9)}&=&\a \sum_{I=1}^4 \phi_I^{(4)}-   
\b \sum_{J=1}^{7}\phi_J^{(2)},\cr   
\cr   
B_0^{(1,5)}&=&-(\a/2)\ \phi_1^{(4)}-\b\ \phi_1^{(2)},\cr   
\cr   
B_0^{(3,5)}&=&-(\a/2) \phi_1^{(4)}+\b\ \phi_1^{(2)},   
\eeqa  
where  
\beqa  
\phi_1^{(4)}&\to& \phi_1^{(4)}+4(\ep_{1,9}+\ep_{3,9})/\a-2(\ep_{1,5}+   
\ep_{3,5})/\a,\cr   
\phi_I^{(4)}&\to& \phi_I^{(4)}-(\ep_{1,9}+\ep_{3,9})/\a,   
\qquad   
I=2,\cdots 4,\cr   
\phi_1^{(2)}&\to& \phi_1^{(2)}+2(\ep_{1,9}-\ep_{3,9})/\b-2(\ep_{1,5}-   
\ep_{3,5})/\b,\cr   
\phi_J^{(2)}&\to& \phi_J^{(2)},\qquad\qquad   
\qquad\qquad   
J=2,\cdots 7.   
\label{finalz4b4}   
\eeqa   
   
\subsection{Green-Schwarz Terms from Non-Planar Diagrams}   
   
In the field theory framework the cancelation procedure of    
the six dimensional chiral anomalies involves one-loop    
and tree-level diagrams. The former generates the anomaly    
polynomial, while the latter cancels it through the    
generalized Green-Schwarz mechanism. This cancelation    
has a simple interpretation from the string theory point of    
view. Consider 1-loop open string diagrams whose low-energy    
limit reproduces the 1-loop field theory anomalies.    
The factorized form of the anomaly polynomial    
comes from the appropriate non-planar diagrams in  
string theory.  
Let $t$ denotes the standard    
1-loop cylinder modulus, the field theory result comes from the    
$t\to +\infty$ boundary of the moduli space. Since    
the non-planar diagrams are finite \cite{GSW}, they can not generate    
chiral anomalies: the anomalous contribution of the 1-loop    
amplitude from the large $t$ region of the moduli space    
is canceled exactly by the contribution from the small    
$t$ region. Through the conformal transformation    
$t\to 1/\ell$, the $t\to 0$ region of the    
1-loop open  string amplitude is interpreted as    
$\ell\to\infty$ limit of a tree-level closed   
string exchange between two boundary states.    
This is precisely the string theory picture of the Green-Schwarz    
mechanism. In the rest of this section, we identify    
the string diagrams whose $t\to \infty$ limit    
produces purely gauge anomalies of the $T^4/\mbz_2$ orientifold    
models. For illustrative purposes, we consider terms of the anomaly    
polynomial (\ref{z2anomaly}) in the ${\mathbb Z}_2$ models
which are sensitive to the    
$2^{b/2}$ multiplicity of states in the 59-open string sector.   
The analysis 
for the other terms in (\ref{z2anomaly}) 
as well as that for the $T^4/\mbz_4$ orientifold    
models is similar and so will not be repeated here.
  
 From the anomaly polynomial (\ref{z2anomaly}),    
the ratio of the coefficients of $f_9\wedge f_9^3$ and    
$f_5\wedge f_9^3$ terms is $(-4/2^{b/2})$. This is    
the result we want to confirm by a string computation.    
   
\begin{figure}   
\centering   
\epsfxsize=6.2in   
\hspace*{0in}\vspace*{.2in}   
\epsffile{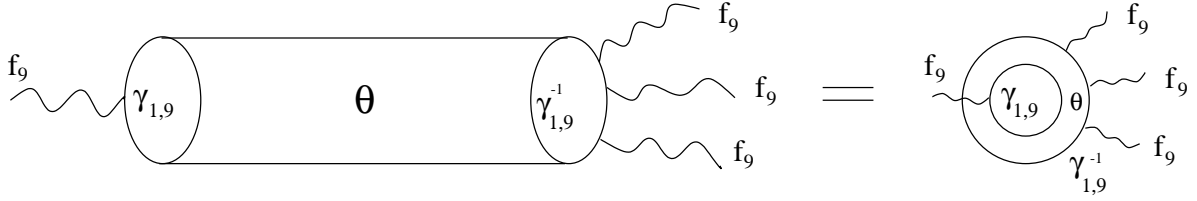}   
\caption{This non-planar digram of the 99-open    
string sector generates the $f_9\wedge f_9^3$ purely    
gauge anomaly in the low-energy limit. Four abelian gauge boson vertex    
operators (represented by $f_9$) are inserted at    
different boundaries of the cylinder. $\g_{1,9}$   
represents  a gauge twist associated with    
the insertion of an orbifold projector $\th$.}   
\label{6dfigure1}   
\end{figure}

We start with the $f_9\wedge f_9^3$ term.  The 1-loop open string    
diagram that would generate this term is shown on    
Fig.~\ref{6dfigure1}.  This amplitude is given by   
\beqa   
A_{(\mu_1;\nu_1\nu_2\nu_3)}^{(99)}\bigg[f_9;f_9^3\bigg]&=&\int 
{d t\over 2t}\    
\biggl[\int dx_1\prod_{n=1}^3   
dy_n\biggr]\cr   
&&\times \biggl\langle\hat{V}_{\mu_1,9}(x_1)   
\hat{V}_{\nu_1,9}(y_1)\hat{V}_{\nu_2,9}(y_2)   
\hat{V}_{\nu_3,9}(y_3)\biggr\rangle_{\{\a,\b\}=\{\oh,\oh\}}^{\cc_{99}},   
\label{ab9ab9}   
\eeqa   
where the 4 99-sector abelian gauge bosons are split into groups of three   
and one    
between the two cylinder boundaries. The subscript of the correlator    
indicates the relevant spin structure\footnote{  
All the vertex operators   
in (\ref{ab9ab9}) are chosen to be in the $(0)$-picture.  
This can be achieved by inserting an appropriate number of   
worldsheet supercurrents.}.  
The correlation function in (\ref{ab9ab9}) decomposes into    
the non-compact part which depends on  the vertex insertion    
coordinates $x_1,y_n$   
but is independent of the twists $\th^k$, and the compact part.   
Since the vertex insertions involve only the non-compact excitations,    
this compact part is identical to the compact part of the $\cc_{99}$   
tadpole with $\{\oh,\oh\}$  spin structure (\ref{z991}). Altogether,    
we can  rewrite  (\ref{ab9ab9}) as   
\beqa   
A_{(\mu_1;\nu_1\nu_2\nu_3)}^{(99)}\bigg[f_9;f_9^3\bigg]   
&=&{1\over 3}\int {d t\over 2t}\  C_{(\mu_1;\nu_1\nu_2\nu_3)}(t)\cr   
&&\times \biggl\{\  \sum_{k=0}^1 \prod_{i=1}^2\    
(-2\sin\pi k v_i)\ \tr(\g_{k,9}\cdot\lambda_9)\tr(\g_{k,9}^{-1}\cdot   
\lambda_9^3)   
\times S_{\cc_{99}}(k) \biggr\}\cr   
&=&{1\over 3}\int {d t\over 2t}\   
C_{(\mu_1;\nu_1\nu_2\nu_3)}(t)\ \biggl\{\ -4\    
\tr(\g_{1,9}\cdot\lambda_9)\tr(\g_{1,9}^{-1}\cdot   
\lambda_9^3)   
\biggr\},   
\label{ab9ab9f}   
\eeqa   
where  $\lambda_9$ is    
the generator of the $U(1)$ in the 99-sector. A symmetry  factor    
$1/3$ in (\ref{ab9ab9f}) accounts for the fact that all three gauge    
bosons on one of the boundaries of the cylinder come from the same $U(1)$.   
   
\begin{figure}   
\centering   
\epsfxsize=6.2in   
\hspace*{0in}\vspace*{.2in}   
\epsffile{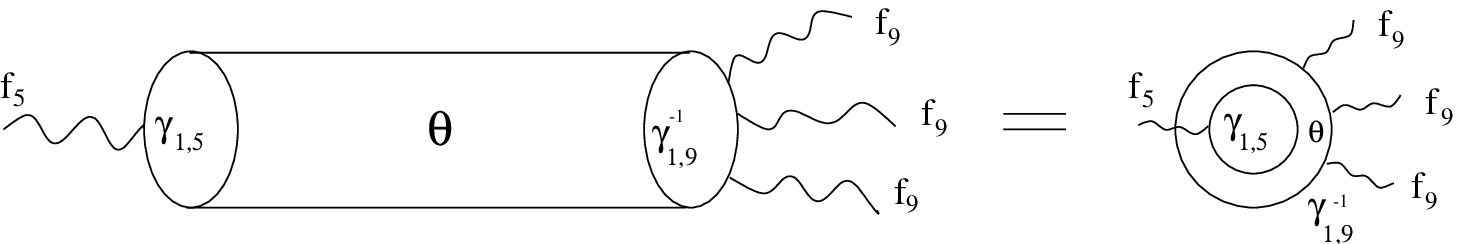}   
\caption{This non-planar digram of the 59-open    
string sector generates the $f_5\wedge f_9^3$ purely    
gauge anomaly in the low-energy limit. Four abelian gauge boson vertex    
operators (represented by $f_5$ and $f_9$) are inserted at    
different boundaries of the cylinder. $\g_{1,9}^{-1}$ and $\g_{1,5}$   
represent  gauge twists associated with    
the insertion of an orbifold projector $\th$.}   
\label{6dfigure2}   
\end{figure}

The 1-loop open string diagram that  generates    
$f_5\wedge f_9^3$ term in the anomaly polynomial    
(\ref{z2anomaly}) is  shown on    
Fig.~\ref{6dfigure2}.  This amplitude is given by   
\beqa   
A_{(\mu_1;\nu_1\nu_2\nu_3)}^{(59)}\bigg[f_5;f_9^3\bigg]&=&   
2^{b/2}\int {d t\over 2t}\    
\biggl[\int dx_1\prod_{n=1}^3   
dy_n\biggr]\cr   
&&\times \biggl\langle\hat{V}_{\mu_1,5}(x_1)   
\hat{V}_{\nu_1,9}(y_1)\hat{V}_{\nu_2,9}(y_2)   
\hat{V}_{\nu_3,9}(y_3)\biggr\rangle_{\{\a,\b\}=\{\oh,\oh\}}^{\cc_{59}},   
\label{ab5ab9}   
\eeqa    
where $2^{b/2}$ accounts for the multiplicity    
of 59-cylinders.   
As before, the correlation function in (\ref{ab5ab9}) decomposes into    
the non-compact part which depends on  the vertex insertion    
coordinates $x_1,y_n$   
but is independent of the twists $\th^k$, and the compact part.   
The non-compact part of this  correlation function    
is identical to the non-compact part of the correlation    
function in  (\ref{ab9ab9}) since 99- and 59-cylinders    
have the same boundary conditions for the non-compact    
excitations. The compact part coincides with the compact part    
of the $\cc_{59}$ tadpole with $\{\oh,\oh\}$  spin structure    
(\ref{z591}). Altogether,    
the  can  rewrite  (\ref{ab5ab9}) as   
\beqa   
A_{(\mu_1;\nu_1\nu_2\nu_3)}^{(59)}\bigg[f_5;f_9^3\bigg]   
&=&{2^{b/2}\over 3}\int {d t\over 2t}\  C_{(\mu_1;\nu_1\nu_2\nu_3)}(t)\cr   
&&\times \biggl\{\  \sum_{k=0}^1 \prod_{i=1}^2\    
  \frac{\vartheta[{{1} \atop {\oh + kv_i}}]}   
{\vartheta[{0 \atop {\oh +kv_i}}]} \   
\tr(\g_{k,5}\cdot\lambda_5)\tr(\g_{k,9}^{-1}\cdot   
\lambda_9^3) \biggr\}\cr   
&=&{2^{b/2}\over 3}\int {d t\over 2t}\   
C_{(\mu_1;\nu_1\nu_2\nu_3)}(t)\ \biggl\{\ \    
\tr(\g_{1,5}\cdot\lambda_5)\tr(\g_{1,9}^{-1}\cdot   
\lambda_9^3)   
\biggr\},   
\label{ab5ab9f}   
\eeqa   
where  $\lambda_5$ is    
the generator of the $U(1)$ in the 55-sector.    
 From (\ref{ab9ab9f}) and (\ref{ab5ab9f})   
\beq   
{A_{(\mu_1;\nu_1\nu_2\nu_3)}^{(99)}\bigg[f_9;f_9^3\bigg]   
\over A_{(\mu_1;\nu_1\nu_2\nu_3)}^{(59)}\bigg[f_5;f_9^3\bigg]}=   
-{4\over 2^{b/2}},   
\label{99ratio59}   
\eeq    
which is the desired result.

\section{Discussion}   
\label{discussion}   
   
In this paper we have considered Type IIB orientifolds    
on $T^4/\mbz_N$ ($N=2,4$) with discrete B-flux, previously    
constructed in \cite{kst9803}.    
We have analyzed the gravitational, gauge and mixed    
anomalies of the resulting six dimensional vacua   
and showed that they all cancel. The cancelation    
required the $2^{b/2}$ multiplicity of states    
in the 59-open string sector. In a field theory    
framework, we have identified the twisted sector R-R    
scalars and tensor multiplets involved in the Green-Schwarz mechanism  
responsible for this cancelation.    
   
We presented details of the construction of these models    
and argued that consistency with $2^{b/2}$ multiplicity    
of 59-sector states requires a modification of the    
relation between the open string 1-loop channel modulus  
and the closed string tree  
channel    
modulus for the 59- (95-) cylinder amplitudes. The latter    
should not be surprising, since only for cylinders with the    
same boundary conditions ( 99- or 55-cylinders) can one   
relate their loop moduli to the loop moduli    
of the Klein bottle and the M\"obius strip by adding/removing    
crosscaps. In fact, we argued that it is precisely    
the anomaly cancelation condition that should be used to determine the    
relation between the 59-cylinder 1-loop and tree channel modulus.   
The reason is that chiral anomalies are sensitive to the    
R-R spin structure of 1-loop amplitudes, however this spin    
structure  does not contribute to the 1-loop tadpoles. On the other 
hand, the R-R   
spin structure will contribute to 2-loop tadpoles,    
and hence the cancelation of anomalies probes the 2-loop tadpole    
consistency of the models.   
   
We showed that the anomaly polynomial   
computed in field theory can be extracted    
from an appropriate limit of non-planar cylinder    
amplitudes. We identified certain 1-loop non-planar    
diagrams which are sensitive to the $2^{b/2}$ multiplicity    
of 59-sector states and showed that they correctly    
reproduce the field theory result, when the multiplicity    
of 59-cylinders is taken to be $2^{b/2}$ as required    
from the cancelation of the tadpoles.   
   
Our results here can be extended to four dimensional orientifolds 
\cite{bst}.  In \cite{ST} a $T^6/\mbz_6$ orientifold with $b=2$ B-flux 
was proposed as an explicit string realization of the ``brane world'' 
scenario.  The model constructed there contains three chiral families 
and a Pati-Salam $SU(4)\otimes SU(2)_L\otimes SU(2)_R$ gauge 
group. The fact that there are three chiral families depends crucially 
on the $2^{b/2}$ multiplicity of states in the 59-open string sector. 
In four dimensions, anomaly constraints are much weaker and so are not 
stringent enough to determine the doubling of the 59-sector states 
\cite{bst}.  However, by compactifying the $b=2$ $T^4/\mbz_2$ 
orientifold considered in this paper on an extra $T^2$ and further 
orbifolding it by $\mbz_3$, we obtain the $T^6/\mbz_6$ orientifold. 
Before the $\mbz_3$ orbifold projection, there is doubling of states 
in the 59 sector of the orientifold since it is connected to the 
six-dimensional $\mbz_2$ orientifold discussed in this paper by 
decompactifying the $T^2$ \footnote{For special radius of the $T^2$, 
it is possible that some momentum/winding states become 
massless. However, these states do not come from the $59$ sector.}. 
This doublet of states can only transform as a singlet representation 
under $\mbz_3$. Thus either all the 59-states of $T^6/\mbz_6$ 
orientifolds are projected out, or all of them are kept.  Since 
tadpole cancelation requires the presence of 59 open string sector 
states \cite{ST,bst}, this implies that the 59 sector states come with 
a multiplicity \makebox{of 2}. In addition to the two anomalous $U(1)$ gauge
fields which contribute to the mixed $U(1)$-gravitational anomalies \cite{ST},
two more anomalous $U(1)$'s arise when other mixed $U(1)$ gauge
anomalies are taken into account, giving rise to a total of 
four anomalous $U(1)$ gauge fields\cite{bst}.
  
The presence of the NS-NS sector B-field background is rather  
generic. For example, when one considers orientifolding an  
asymmetric orbifold \cite{bg9812158}, the background  
B-field is generically non-zero.    
The fact that that $b_{ij}$ defined in Eq.(\ref{momenta}) is invariant  
under $b_{ij} \rightarrow b_{ij}+1$ and that it  
takes only quantized values of $0$ or $1/2$ suggests    
that it should be possible to associate with it a     
$\mbz_2$ action. Indeed, one also finds   
a $2^{b/2}$ multiplicity of states in the 59-sector in this  
picture \cite{bg9812158}.

\acknowledgments   
   
{}We thank Philip Argyres,   
Mirjam Cvetic, Luis Ibanez, Clifford Johnson, Zurab Kakushadze, 
Robert Shrock, Angel Uranga  
and Piljin Yi for discussions. The research of G.S. is partially   
supported by the NSF   
grant PHY-97-22101.   
The research of A.B. and    
S.-H.H.T. is partially supported by   
the National Science Foundation.    
G.S. and S.-H.H.T would also like to thank the Aspen Center for Physics   
for hospitality  
while part of this work was completed.

\appendix 
\label{appendix6d}   
   
\section{One-loop vacuum amplitudes  
in Orientifolds  
without NS-NS antisymmetric tensor background.}   
\label{1loopb0}   
   
In this section we review the computation of 1-loop vacuum amplitudes  
of the $T^4/\mbz_N$ orientifolds \cite{gp,gj9606}.    
The 1-loop vacuum amplitudes come from the Klein bottle $\ck$,    
M\"{o}bius strip $\cam$, and cylinder $\cc$.

The Klein bottle amplitude is given by   
\beq   
\ck = \frac{i V_6}{8N} \sum_{n,k=0}^{N-1}  \int_0^\infty \, \frac{dt}t \,   
(4\pi^2 \a^\prime t)^{-3} \, \cz_{\ck}(\th^n,\th^k),   
\label{kamp}   
\eeq   
where   
\beq   
\cz_{\ck}(\th^n,\th^k) = \Tr \{(1+(-1)^F) \Om \, \th^k \,   
{\rm e}^{-2\pi t[L_0(\th^n) + \ov{L}_0(\th^n)]} \}.   
\label{zkdef}   
\eeq   
The contribution of the uncompactified momenta is already extracted in   
(\ref{kamp}). Here, $V_6$ denotes the regularized space-time volume.   
Since $\Om$ exchanges $\th^n$ with $\th^{N-n}$, only $n=0$ and   
$n=N/2$ terms survive the trace.   
The trace in $\cz_{\ck}$ can be evaluated in a standard way using   
$\vt$ functions to write the contributions of complex bosons and fermions.   
Also, the GSO projection is implemented by summing over spin structures.   
Then, taking into account the insertion of $\Om$ we find   
\beq   
\cz_{\ck}(1,\th^k) = \sum_{\a,\b=0,\oh}  \eta_{\a,\b} \   
\frac{\tilde\vt^2[{\a \atop \b}]}{\tilde\eta^6} \ \prod_{i=1}^2   
\  (-2\sin 2\pi k v_i) \, \frac{\tilde\vt[{\a \atop    
{\b + 2kv_i}}]}{\tilde\vt[{\oh \atop {\oh +2kv_i}}]}    
\times S_{\ck}(k),   
\label{k1}   
\eeq   
where $\eta_{0,0} =-\eta_{0,\oh} = -\eta_{\oh,0} =1$.   
The $\vt$ and $\eta$ functions are defined in  
Appendix~\ref{thetaproperties}.    
The tilde indicates that the argument is   
$\tilde q = q^2 = {\rm e}^{-4\pi t}$. Note that for    
$2kv_i  = {\rm integer}$, (\ref{k1}) has a well-defined limit.    
The factors $S_{\ck}(k)$ account for the sums over quantized momenta    
($kv_i=$integer) or winding ($kv_i=$half-integer) in the $Y_i$ direction:     
\beqa   
S_{\ck}(k)&=&1,\qquad\qquad\qquad\qquad\qquad\qquad\qquad\qquad\quad   
 k\ne\{0,N/2\},\cr   
\cr   
S_{\ck}(k)&=&\delta_{k,0}\sum_{p^2\in \star\Gamma_4}q^{\alpha' p^2/2}+   
\delta_{k,N/2}\sum_{\om^2\in \Gamma_4}q^{\om^2/(2 \alpha')},\qquad    
k=\{0,N/2\}.   
\label{skb}   
\eeqa   
In (\ref{skb}), $\Gamma_4$ and $\star\Gamma_4$ denote the    
winding and momentum lattice of $T^4$ correspondingly.   
The $t\to 0$ divergence of (\ref{k1}) represents the exchange of the    
massless NS-NS (terms with $\{\a,\b\}=\{0,0\}$ and $\{\a,\b\}=\{\oh,0\}$   
in (\ref{k1})) and    
R-R forms in the closed string tree channel.  
    
{}The $Z_2$ twisted sector yields   
\beq   
\cz_{\ck}(\th^{N/2},\th^k) =  \tilde \chi(\th^{N/2}, \th^k)     
\sum_{\a,\b=0,\oh} \! \eta_{\a,\b} \   
\frac{\tilde\vt^2[{\a \atop \b}]}{\tilde\eta^6}  \prod_{i=1}^2   
\frac{\tilde\vt[{{\a+\oh} \atop {\b + 2kv_i}}]}   
{\tilde\vt[{0 \atop {\oh +2kv_i}}]}.    
\label{kR}   
\eeq   
In eq.~(\ref{kR}), $\tilde \chi(\th^{N/2}, \th^k)$ is a factor that
takes into account the fixed point degeneracy \cite{orbi}, {\em i.e.},
it counts the number of fixed points of $\th^{N/2}$ on $T^4$ invariant
under $\th^k$:
\beqa   
\tilde \chi(\th^{N/2}, \th^k)&=&16,\qquad k=\{0,N/2\},\cr   
\tilde \chi(\th^{2}, \th^k)&=&4,\qquad k=\{1,3\},\ {\rm and}\  N=4.   
\label{z2fixed}   
\eeqa   
As (\ref{k1}),  (\ref{kR}) vanishes by virtue of supersymmetry.   
The NS-NS exchange in the closed string tree channel is represented by    
$\{\a,\b\}=\{0,0\}$ and $\{\a,\b\}=\{\oh,0\}$ terms in (\ref{kR}).   
   
The cylinder amplitudes are given by   
\beq   
\cc_{pq} = \frac{iV_6}{8N} \sum_{k=0}^{N-1}  \int_0^\infty \, \frac{dt}t \,   
(8\pi^2 \a^\prime t)^{-3} \, \cz_{pq}(\th^k),   
\label{camp}   
\eeq   
where   
\beq   
\cz_{pq}(\th^k) = \Tr_{pq} \{(1+(-1)^F)  \th^k \,   
{\rm e}^{-2\pi t L_0} \}.   
\label{zcdef}   
\eeq   
The trace is over open string states with boundary conditions according   
to the D$p$ and D$q$-branes at the endpoints.    
   
{}In $\cz_{99}$, boundary conditions are NN   
in all directions. Hence,   
\beq   
\cz_{99}(\th^k) = \sum_{\a,\b=0,\oh}  \eta_{\a,\b} \   
\frac{\vartheta^2[{\a \atop \b}]}{\eta^6} \ \prod_{i=1}^2   
\ (-2\sin \pi k v_i) \,    
\frac{\vartheta[{\a \atop {\b + kv_i}}]}{\vartheta[{\oh \atop 
{\oh +kv_i}}]}\   
(\Tr\g_{k,9})^2\ \times S_{\cc_{99}}(k).   
\label{z991}   
\eeq   
The factors $S_{\cc_{99}}(k)$ account for the sums over quantized momenta    
in $Y_i$ when $kv_i=$integer :   
\beqa   
S_{\cc_{99}}(k)&=&1,\qquad\qquad   
 k\ne 0,\cr   
\cr   
S_{\cc_{99}}(0)&=&\sum_{p^2\in \star\Gamma_4}q^{\alpha' p^2}.   
\label{scyl}   
\eeqa   
Eq.~(\ref{z991}) vanishes by supersymmetry. The terms $\{\a,\b\}=\{0,0\}$   
and $\{\a,\b\}=\{\oh,0\}$ represent NS-NS exchange in the    
closed string tree channel.    
   
In the 55-sector there are DD boundary conditions in directions $Y_1,Y_2$   
transverse to the 5-branes.    
The oscillator expansions with DD boundary conditions   
have integer modes but include windings instead of momenta. Then, $\cz_{55}$   
has a form similar to (\ref{z991})   
\beq   
\cz_{55}(\th^k) = \sum_{\a,\b=0,\oh}  \eta_{\a,\b} \   
\frac{\vartheta^2[{\a \atop \b}]}{\eta^6}\ \prod_{i=1}^2   
\ (-2\sin \pi k v_i) \,    
\frac{\vartheta[{\a \atop {\b + kv_i}}]}{\vartheta[{\oh \atop {\oh +kv_i}}]}   
\ \sum_{I=1}^{n(N,k)} \, (\Tr\g_{k,5,I})^2\ \times S_{\cc_{55}}(k),   
\label{z55}   
\eeq   
where $I$ refers to the fixed points of $\th^k$. Thus,   
\beqa   
&&n(2,1)=n(4,2)=16,\cr   
&&n(4,1)=n(4,3)=4.   
\label{55fixed}   
\eeqa    
$S_{\cc_{55}}(k)$ account for the sums over winding    
in $Y_i$ when $kv_i=$integer :   
\beqa   
S_{\cc_{55}}(k)&=&1,\qquad\qquad   
 k\ne 0,\cr   
\cr   
S_{\cc_{55}}(0)&=&\sum_{\om^2\in \Gamma_4}q^{\om^2/\alpha' }.   
\label{scyl55}   
\eeqa   
In (\ref{z55}) terms $\{\a,\b\}=\{0,0\}$ and $\{\a,\b\}=\{\oh,0\}$   
represent NS-NS exchange in the closed string tree channel.   
   
{}In the 59-sector there are DN boundary conditions in    
coordinates $Y_1,Y_2$.   
Hence, their oscillator expansions include half-integer modes. For fermions,   
world-sheet supersymmetry requires  that in Neveu-Schwarz (Ramond),   
the moddings   
are opposite (same) to that of the corresponding bosons. Hence,   
\beq   
\cz_{59}(\th^k) = \sum_{\a,\b=0,\oh}  \eta_{\a,\b} \,   
\frac{\vartheta^2[{\a \atop \b}]}{\eta^6}   
\, \prod_{i=1}^2   
\  \frac{\vartheta[{{\a+\oh} \atop {\b + kv_i}}]}   
{\vartheta[{0 \atop {\oh +kv_i}}]} \,   
\Tr\g_{k,9} \sum_I^{n(N,k)} \, \Tr\g_{k,5,I}.   
\label{z591}   
\eeq   
The $\{\a,\b\}=\{0,0\}$ and $\{\a,\b\}=   
\{\oh,0\}$ contributions of (\ref{z591}) represent the    
NS-NS exchange in the closed string tree channel.    
   
The M\"obius strip amplitudes are given by   
\beq   
\cam_p = \frac{i V_6}{8N} \sum_{k=0}^{N-1}  \int_0^\infty \, \frac{dt}t \,   
(8\pi^2 \a^\prime t)^{-3} \, \cz_p(\th^k),   
\label{mamp}   
\eeq   
where   
\beq   
\cz_p(\th^k) = \Tr_p \{(1+(-1)^F) \, \Om \, \th^k \,   
{\rm e}^{-2\pi t L_0} \}.   
\label{zmdef}   
\eeq   
Here the trace is over open string states with boundary conditions   
according to the D$p$-branes at both endpoints. The main difference   
between $\cz_p$ and $\cz_{pp}$ is the insertion of $\Omega$ that   
acts on the various bosonic and fermionic oscillators thereby   
introducing extra phases in the expansions in $q$. More precisely,   
$\Omega$ acts on oscillators as \cite{gp}   
\beq   
\a_r \to  \pm {\rm e}^{i\pi r}   
\quad\quad ; \quad\quad   
\psi_r \to  \pm {\rm e}^{i\pi r}.   
\label{apsi}   
\eeq   
The upper (lower) sign is for NN (DD) boundary conditions.   
Furthermore, $\Omega$   
acts as ${\rm e}^{-i\pi/2}$ on the NS vacuum. This ensures that   
$\Om( \psi^\mu_{-\oh} \vac_{NS})  =  - \psi^\mu_{-\oh} \vac_{NS}$ as needed   
for the orientifold projection on gauge vectors.   
   
{}To derive the M\"obius trace we can use (\ref{apsi}) and the results for   
99-cylinders. After using $\vt$ identities we obtain   
\beqa   
\cz_9(\th^k) &=& -(1-1) \,   
\frac {\tilde\vt^2[{\oh \atop 0}]\, \tilde\vt^2[{0 \atop \oh }]}   
{\tilde\eta^6 \tilde\vt^2[{0 \atop 0}]} \ \prod_{i=1}^2 \,   
\frac{-2\sin \pi k v_i \,   
\tilde\vt[{\oh \atop {kv_i}}]\, \tilde\vt[{0 \atop {\oh + kv_i}}]}   
{\tilde\vt[{\oh \atop {\oh + kv_i}}] \, \tilde\vt[{0 \atop {kv_i}}]}   
\, \Tr(\g_{\Omega_k,9}^{-1} \cdot \g_{\Omega_k,9}^T)\cr   
&& \times S_{\cam_{9}}(k).   
\label{zm9}   
\eeqa   
In (\ref{zm9}) we separate the NS-NS and R-R exchange in the    
closed string tree channel. $S_{\cam_{9}}(k)$  account for   
a sum over quantized momentum in $Y_i$ for $kv_i=$integer:   
\beqa   
S_{\cam_{9}}(k)&=&1,\qquad\qquad   
 k\ne 0,\cr   
\cr   
S_{\cam_{9}}(0)&=&\sum_{p^2\in \star\Gamma_4}q^{\alpha' p^2}.   
\label{sm9}   
\eeqa

{}For $5$-branes, we find   
\beqa   
\cz_5(\th^k) & = & (1-1) \,   
\frac {\tilde\vt^2[{\oh \atop 0}]\, \tilde\vt^2[{0 \atop \oh }]}   
{\tilde\eta^6 \tilde\vt^2[{0 \atop 0}]} \,   
 \prod_{i=1}^2 \,   
\frac{2\cos \pi k v_i \,   
\tilde\vt[{\oh \atop {\oh + kv_i}}]\, \tilde\vt[{0 \atop kv_i}]}   
{\tilde\vt[{\oh \atop kv_i}] \, \tilde\vt[{0 \atop {\oh + kv_i}}]}   
\, \sum_I \Tr(\g_{\Omega_k,5,I}^{-1} \cdot \g_{\Omega_k,5,I}^T)\cr   
&& \times S_{\cam_{5}}(k).   
\label{zm5}   
\eeqa   
In (\ref{zm5}) we separate the NS-NS and R-R exchange in the    
closed string tree channel.   
$S_{\cam_{5}}(k)$  account for   
a sum over winding in $Y_i$ for $kv_i=$half-integer:   
\beqa   
S_{\cam_{5}}(k)&=&1,\qquad\qquad   
 k\ne N/2,\cr   
\cr   
S_{\cam_{5}}(N/2)&=&\sum_{\om^2\in \Gamma_4}q^{\om^2/\alpha' }.  
\label{sm5}   
\eeqa

\section{Some properties of the $\vt$ functions}   
\label{thetaproperties}   
   
The $\vt$ function of rational characteristics $\d$ and $\vphi$ is given by   
\beq   
\vt[{\d \atop \vphi}](t) = \sum_n q^{\oh(n+\d)^2} \,   
{\rm e}^{2i\pi (n+\d) \vphi}.   
\label{vts}   
\eeq   
Here the variable $q$ is $q={\rm e}^{-2\pi t}$.   
The $\vt$ function also has the product form   
\beq   
\frac{\vt[{\d \atop \vphi}]}{\eta} = {\rm e}^{2i\pi \d \vphi}   
\, q^{\oh \d^2 - \frac1{24}} \,   
\prod_{n=1}^\infty (1 + q^{n+\d -\oh} {\rm e}^{2i\pi \vphi} ) \,   
(1 + q^{n-\d -\oh} {\rm e}^{-2i\pi \vphi} ),   
\label{vtp}   
\eeq   
where the Dedekind $\eta$ function is   
\beq   
\eta =  q^{\frac1{24}} \,   
\prod_{n=1}^\infty (1 - q^n).   
\label{deta}   
\eeq   
Notice that   
\beq   
\lim_{\vphi \to 0} \,   
\frac{-2 \sin \pi \vphi }{\vt[{\oh \atop {\oh+\vphi}}]} =   
\frac1{\eta^3}.   
\label{limu}   
\eeq   
The $\vt$ and $\eta$ functions have the modular transformation properties   
\beqa   
\vt[{\d \atop \vphi}](t) & = &{\rm e}^{2i\pi \d \vphi} \, t^{-\oh} \,   
\vt[{-\vphi \atop \d}](1/t),   
\nonumber \\[0.2ex]   
\eta(t) & = &  t^{-\oh} \, \eta(1/t).   
\label{modt}   
\eeqa   
The $\vt$'s satisfy several Riemann identities \cite{mum}. In particular,   
\beqa   
& {} & \sum_{\a,\b} \eta_{\a,\b} \  \vt[{\a \atop \b}] \prod_{i=1}^3   
\vt[{\a \atop {\b+u_i}}] = 0,   
\nonumber \\[0.2ex]   
& {} & \sum_{\a,\b} \eta_{\a,\b} \ \vt[{\a \atop \b}]   
\vt[{\a \atop {\b+u_3}}]   
\prod_{i=1}^2 \vt[{{\a+\oh} \atop {\b+u_i}}] = 0,   
\label{abtru}   
\eeqa   
provided that $u_1+u_2+u_3=0$.

\section{Chiral anomalies in six dimensions}   
\label{anomalyreview}   
In this section we collect some facts about anomalies in six    
dimensional $\cn=1$ supergravity theories. Details can be   
found in Ref. \cite{alvarez,gswest,erler,sreview}.   
   
$\cn=1$ supersymmetry in six dimensions involves four types of    
massless multiplets. In terms of representations of    
the little group $SU(2)\otimes SU(2)$ (we label the $SU(2)$    
representations by their multiplicities $(2J+1)$), the massless  
multiplets are   
   
\noindent  (i) gravity multiplet  
$G$: ${\bf(3,3)}\oplus 2{\bf(3,2)}\oplus {\bf(3,1)}$,   
   
\noindent  (ii) tensor multiplet $T$:   
${\bf(1,3)}\oplus 2{\bf(1,2)}\oplus {\bf(1,1)}$,   
   
\noindent  (iii) vector multiplet $V$:   
${\bf(2,2)}\oplus 2{\bf(2,1)}$,     
   
\noindent  (iv)  hypermultiplet $H$:    
$2{\bf(1,2)}\oplus 4{\bf(1,1)}$.   
    
In six dimensions    
chiral anomalies are characterized by an 8-form polynomial    
constructed from the gauge and gravitational field strengths.   
The basic diagram to be examined is the box diagram with an even    
number of external gravitons and gauge fields. The resulting    
anomalous diagrams can be classified as purely gravitational,   
purely gauge or mixed gauge and gravitational.     
   
All four $\cn=1$ supergravity multiplets contribute to purely     
gravitational anomaly. Their separate contributions to the    
anomaly polynomial are given by   
\beqa   
i{(2\pi)}^3 I_G^{grav.}&=&{1\over 5760} \biggl(273\ \tr R^4-{5\over 4}\ 51    
\ {(\tr R^2)}^2\biggr),\cr   
\cr   
i{(2\pi)}^3 I_T^{grav.}&=&{1\over 5760} \biggl(-29\ \tr R^4+{5\over 4}\ 7    
\ {(\tr R^2)}^2\biggr),\cr   
\cr   
i{(2\pi)}^3 I_V^{grav.}&=&{1\over 5760} \biggl( \tr R^4+{5\over 4}    
\ {(\tr R^2)}^2\biggr),\cr   
\cr   
i{(2\pi)}^3 I_H^{grav.}&=&{1\over 5760} \biggl( -\tr R^4-{5\over 4}    
\ {(\tr R^2)}^2\biggr).   
\label{grav}   
\eeqa     
where the trace over curvature matrices in $R$ is in the    
vector representation of $SO(5,1)$. The cancelation of 
the leading term $\tr R^4$ in the pure gravitational anomalies implies 
\begin{equation} 
n_H - n_V = 273 - 29 n_T 
\end{equation} 
   
Mixed gauge and gravitational anomalies get contributions    
from gauginos and matter fermions:   
\beqa   
i{(2\pi)}^3 I^{mixed}=-{1\over 96}\ \tr R^2    
\biggl(&&\sum_A\Tr F^2_A-\sum_{I,A}s_A^I(\tr_{R^I}F^2_A)\cr   
&&-\sum_{i,a}s_a^i\ q_{i,a}^2\ (\tr f^2_a)   
-2\sum_{ij,a<b}s_{ab}^{ij}\ q_{ij,a}\ q_{ij,b}\ (\tr f_a)\ 
(\tr f_b)\biggr),    
\label{mixed}   
\eeqa   
where $F_A$ represents the nonabelian field strength of a gauge group
factor $G_A$ and $f_a$ represents the abelian field strength. $s_A^I$
denotes the number of hypermultiplets transforming in representation
$R^I$ of the nonabelian factor $G_A$, while $s_a^i$ counts the
hypermultiplets with $q_{i,a}$ charge under $U(1)_a$.  $s_{ab}^{ij}$
is a multiplicity of states with charges $(q_{ij,a},q_{ij,b})$ under
$U(1)_a\otimes U(1)_b$. $\Tr$ refers to the trace in the respective
adjoint representation.
   
Finally, purely gauge anomaly is given by
\beqa   
i{(2\pi)}^3 I^{gauge}&=&{1\over 24}\ \biggl(\sum_A\Tr F_A^4-\sum_{I,A}s_A^I    
(\tr_{R^I} F_A^4)-6\sum_{IJ,A<B} s_{AB}^{IJ}\ (\tr_{R^I}F_A^2)\    
(\tr_{R^J}F_B^2 )\cr   
&&-4\sum_{Ij,Aa} s_{Aa}^{Ij}\ q_{Ij,a}\ (\tr_{R^I}F_A^3)\ (\tr f_a)\cr   
&&-6\sum_{Ij,Aa} s_{Aa}^{Ij}\ q_{Ij,a}^2\ (\tr_{R^I}F_A^2)\ {(\tr f_a)}^2\cr   
&&-12\sum_{Iij,A,a<b} s_{Aab}^{Iij}\ q_{Iij,a}\ q_{Iij,b}\ (\tr_{R^I}F_A^2)\   
(\tr f_a)\ (\tr f_b) \cr   
&&-\sum_{i,a}s_a^i\ q_{i,a}^4\ {(\tr f_a)}^4-4\sum_{ij,ab}s^{ij}_{ab}\   
q_{ij,a}^3\ q_{ij,b}\ {(\tr f_a)}^3\ (\tr f_b)\cr   
&&-6\sum_{ij,a<b} s^{ij}_{ab}\   
q_{ij,a}^2\ q_{ij,b}^2\ {(\tr f_a)}^2\ {(\tr f_b)}^2 \biggr),   
\label{gauge}   
\eeqa   
where $s_{AB}^{IJ}$ is the multiplicity of states in the $(R^I,R^J)$   
representation of $G_A\otimes G_B$, $s_{Aa}^{Ij}$ is the multiplicity   
of states in the representation/charge $(R^I,q_{Ij,a})$ of    
$G_A\otimes U(1)_a$, and $s_{Aab}^{Iij}$ is the multiplicity    
of states in the representation/charges $(R^I,q_{Iij,a},q_{Iij,b})$   
of $G_A\otimes U(1)_a\otimes U(1)_b$.   
In (\ref{gauge}) we included only hypermultiplet representations   
which are relevant to the orientifold models of interest    
(see Table \ref{6dtable1} for the hypermultiplet spectrum).   
   
It is    
convenient to express all traces over  nonabelian field strengths    
in the fundamental representation of  respective gauge groups.   
Here we list out the relations between traces in different    
representations for $SU(n)$ groups \cite{erler} (since  
they are the  
nonabelian gauge    
groups arise in the Type IIB orientifolds of interest in this paper).  
For $n \ge 3$,   
\beqa   
\Tr F^4&=&2n\ \tr F^4 + 6\ {(\tr F^2)}^2,\cr   
\tr_{a^{ij}} F^4&=&(n-8)\ \tr F^4 + 3\ {(\tr F^2)}^2,\cr   
\tr_{a^{ij}} F^3&=&(n-4)\ \tr F^3,\cr   
\Tr F^2&=&2n\ \tr F^2,\cr   
\tr_{a^{ij}} F^2&=&(n-2)\ \tr F^2,   
\label{tracesrelations}   
\eeqa    
where $a^{ij}$ denotes a second rank antisymmetric tensor representation.   
$\tr$ refers to the trace in the fundamental representation.   
Furthermore, for $SU(2)$:   
\beqa   
\tr F^4&=&{\oh}\ {(\tr F^2)}^2.   
\label{su2traces}   
\eeqa    
   
\section{Traces over gauge group generators.}   
\label{lambdatraces}   
In Sec.~\ref{opensector} we discuss how the ``shift'' vector    
formalism of \cite{ibanez} can be used to represent gauge twists $\g_{1,a}$   
and determine massless states in the open string sector of the    
Type IIB orientifold models. Here, we collect the formulas relevant    
to the computation of traces for the anomaly cancelation.   
   
In Sec.~\ref{opensector} we rewrite the  
Chan-Paton matrices $\lambda_A^p$ and    
$\lambda_a^p$  in a Cartan-Weyl basis.   
The subscript $A$ refers to the    
nonabelian gauge group factor $G_A$ in the $p=5,9$ open string sector,   
while $a$ labels the generator of $U(1)_a$ in open string sector $p$.   
The Cartan generators are represented    
by tensor product of $2\times 2$ $\sigma_3$ submatricies. We chose    
the normalization of the $SO(32)$ generators $\lambda$ in such    
a way that $\Tr\lambda^2=1$. In all the models    
we discuss, there is a $U(1)$ factor for each of the    
$SU(n)$'s in the model. The generator $\lambda_a^p$    
of a given $U(1)_a$ is given by a linear combination    
\beq   
\lambda_a^p=Q_a^p\cdot H,   
\label{u1gen}   
\eeq         
where $H$ is a vector of Cartan generators    
and $Q_a^p$ is a $16$-dimensional real vector    
of the form    
\beq   
Q_a^p={1\over \sqrt{2 n_a}}(0,0,\cdots,\     
\underbrace{1,\cdots,1}_{n_a\ {\rm times}},\ 0,0,\cdots,0 ),   
\label{chargevector}   
\eeq       
where the non-zero entries sit at the positions where the corresponding   
$U(n_a)$ lives. With these convention and normalization   
\beqa   
&&\tr(\g_{k,p}\ \lambda_a^p)=\tr(e^{-2i\pi k V_{(pp)}\cdot H}\ Q_a^p\cdot H)   
= (-i)\sqrt{2n_a}\ \sin 2\pi k V_{(pp)}^a,\cr   
&&\tr(\g_{k,p}\ \lambda_A^p)=0,\cr   
&&\tr(\g_{k,p}\ {(\lambda_a^p)}^2)=\cos 2\pi k V_{(pp)}^a,\cr   
&&\tr(\g_{k,p}\ {(\lambda_A^p)}^2)=\cos 2\pi k V_{(pp)}^A,\cr   
&&\tr(\g_{k,p}\ {(\lambda_a^p)}^3)=(-i){1\over\sqrt{2n_a}}\    
\sin 2\pi k V_{(pp)}^a,\cr   
&&\tr(\g_{k,p}\ {(\lambda_A^p)}^3)=(-i){1\over\sqrt{2}}\    
\sin 2\pi k V_{(pp)}^A,\cr   
&&\tr(\g_{k,p}\ \lambda_a^p\ {(\lambda_A^p)}^2)=(-i){1\over\sqrt{2n_a}}\    
\sin 2\pi k V_{(pp)}^A,\qquad {\rm when}\ V_{(pp)}^a=V_{(pp)}^A.    
\label{tracescollection}   
\eeqa   
where $n_a$ is the rank of the $U(n)$ group containing  $U(1)_a$ and    
$V_{(pp)}^a$ is a component of the $V_{(pp)}$ shift vector along    
any of the overlapping entries with that $U(n)$, $V_{(pp)}^A$    
is a component of  $V_{(pp)}$ along any entry  overlapping  with the group   
$G_A$ in the $p$-open string sector.

\begin{table}[t]   
\begin{tabular}{|c|c|c|l|c|c|}   
Model & $b$ & Gauge Group & Charged  &    
Neutral  &  Extra Tensor\\   
& & & Hypermultiplets & Hypermultiplets &  Multiplets \\    
\hline   
$\mbz_2$& 0& $[SU(16)\otimes U(1)]^2$&$2\times ({\bf 120;1})(+1/2;0)$   
&20&0\\   
& & & $2\times ({\bf 1; 120})(0;+1/2)$ & & \\   
& & &  $({\bf 16; \overline{16}})(+1/4;-1/4)$ & & \\   
\hline   
$\mbz_2$& 2& $[SU(8)\otimes U(1)]^2$&$2\times ({\bf 28;1})(+1/\sqrt{2};0)$   
&16&4\\   
& & & $2\times ({\bf 1; 28})(0;+1/\sqrt{2})$ & & \\   
& & &  $2\times ({\bf 8; \overline{8}})(+1/\sqrt{8};-1/\sqrt{8})$ & & \\   
\hline   
$\mbz_2$& 4& $[SU(4)\otimes U(1)]^2$&$2\times ({\bf 6;1})(+1;0)$   
&14&6\\   
& & & $2\times ({\bf 1; 6})(0;+1)$ & & \\   
& & &  $4\times ({\bf 4; \overline{4}})(+1/2;-1/2)$ & & \\   
\hline   
$\mbz_4$& 0& $[SU(8)\otimes SU(8)$&   
$({\bf 28,1;1,1})(+1/\sqrt{2},0;0,0)$ &16&4\\   
& & $\otimes U(1)^2]^2$&    
$({\bf 1,\overline{28}; 1,1})(0,-1/\sqrt{2};0,0)$ & & \\   
& & & $({\bf 1,1; 28,1})(0,0;+1/\sqrt{2},0)$ & &\\   
& & & $({\bf 1,1; 1, \overline{28}})(0,0;0,-1/\sqrt{2})$ & & \\   
& & & $({\bf 8,\overline{8};1,1})(+1/\sqrt{8},-1/\sqrt{8};0,0)$ & & \\   
& & & $({\bf 1,1;8,\overline{8}})(0,0;+1/\sqrt{8},-1/\sqrt{8})$ & & \\   
& & & $({\bf 8,1;\overline{8},1})(+1/\sqrt{8},0;-1/\sqrt{8},0)$ & & \\   
& & & $({\bf 1,8;1,\overline{8}})(0,+1/\sqrt{8};0,-1/\sqrt{8})$ & & \\   
\hline   
$\mbz_4$& 2& $[SU(4)\otimes SU(4)$&   
$({\bf 6,1;1,1})(+1,0;0,0)$ &14&6\\   
& & $\otimes U(1)^2]^2$&    
$({\bf 1,\overline{6}; 1,1})(0,-1;0,0)$ & & \\   
& & & $({\bf 1,1; 6,1})(0,0;+1,0)$ & &\\   
& & & $({\bf 1,1; 1, \overline{6}})(0,0;0,-1)$ & & \\   
& & & $({\bf 4,\overline{4};1,1})(+1/2,-1/2;0,0)$ & & \\   
& & & $({\bf 1,1;4,\overline{4}})(0,0;+1/2,-1/2)$ & & \\   
& & & $2\times ({\bf 4,1;\overline{4},1})(+1/2,0;-1/2,0)$ & & \\   
& & & $2\times ({\bf 1,4;1,\overline{4}})(0,+1/2;0,-1/2)$ & & \\   
\hline   
$\mbz_4$& 4& $[SU(2)\otimes SU(2)$&   
$({\bf 1,1;1,1})(+\sqrt{2},0;0,0)$ &13&7\\   
& & $\otimes U(1)^2]^2$&    
$({\bf 1,1; 1,1})(0,-\sqrt{2};0,0)$ & & \\   
& & & $({\bf 1,1; 1,1})(0,0;+\sqrt{2},0)$ & &\\   
& & & $({\bf 1,1; 1,1})(0,0;0,-\sqrt{2})$ & & \\   
& & & $({\bf 2,2;1,1})(+1/\sqrt{2},-1/\sqrt{2};0,0)$ & & \\   
& & & $({\bf 1,1;2,2})(0,0;+1/\sqrt{2},-1/\sqrt{2})$ & & \\   
& & & $4\times ({\bf 2,1;2,1})(+1/\sqrt{2},0;-1/\sqrt{2},0)$ & & \\   
& & & $4\times ({\bf 1,2;1,2})(0,+1/\sqrt{2};0,-1/\sqrt{2})$ & & \\   
\hline   
\end{tabular}   
\caption{The massless spectrum of the six dimensional Type IIB
orientifolds on $T^4/\mbz_N$ for $N=2,4$, and various values of $b$
(the rank of $B_{ij}$).  The semi-colon in the column ``Charged
Hypermultiplets'' separates 99 and 55 representations.}
\label{6dtable1}   
\end{table}

\end{document}